 \documentclass[final,5p,times,twocolumn]{elsarticle}

\usepackage{amsmath,amssymb,amsthm,epsfig}
\usepackage{multirow}
\usepackage{threeparttable}
\usepackage{epstopdf}
\usepackage{tikz}
\usepackage{graphicx}
\usetikzlibrary{positioning}

\DeclareSymbolFont{matha}{OML}{txmi}{m}{it}
\DeclareMathSymbol{\varv}{\mathord}{matha}{118}

\journal{Pattern Recognition}

\begin{document}

\begin{frontmatter}



\title{Three-Dimensional Krawtchouk Descriptors for Protein Local Surface Shape Comparison}

\author[label1]{Atilla Sit}
\address[label1]{Department
of Mathematics and Statistics, Eastern Kentucky University, Richmond, KY, 40475 USA}
\ead{atilla.sit@eku.edu}

\author[label2,label3,label4]{Daisuke Kihara\corref{cor1}}
\address[label2]{Department of Biological Sciences, Purdue University, West Lafayette, IN, 47907 USA}
\address[label3]{Department of Computer Science, Purdue University, West Lafayette, IN, 47907 USA}
\address[label4]{Department of Pediatrics,  Cincinnati Children's Hospital Medical Center, University of Cincinnati, Cincinnati, OH, 45229 USA}
\cortext[cor1]{Corresponding author.}
\ead{dkihara@purdue.edu}


\begin{abstract}
Direct comparison of three-dimensional (3D) objects is computationally expensive due to the need for translation, rotation, and scaling of the objects to evaluate their similarity. In applications of 3D object comparison, often identifying specific local regions of objects is of particular interest. We have recently developed a set of 2D moment invariants based on discrete orthogonal Krawtchouk polynomials for comparison of local image patches. In this work, we extend them to 3D and construct 3D Krawtchouk descriptors (3DKD) that are invariant under translation, rotation, and scaling. The new descriptors have the ability to extract local features of a 3D surface from any region-of-interest. This property enables comparison of two arbitrary local surface regions from different 3D objects. We present the new formulation of 3DKD and apply it to the local shape comparison of protein surfaces in order to predict ligand molecules that bind to query proteins.
\end{abstract}


\end{frontmatter}


\section{Introduction}
\label{intro}

Moment-based approaches have become very popular in 2D and 3D image processing due to their compact representation of images. A moment-based approach characterizes a 2D or 3D image by considering its shape as a mathematical function and computes integral of the function multiplied by specific base functions. The approach has been used in many problems including reconstruction, detection, pattern recognition, and compression of images \cite{ShuEtAl2007}. The theory of moment invariants in 2D has been well established since the foundation of algebraic Hu invariants \cite{Hu1962}. Sadjadi and Hall \cite{SadjadiHall1980} extended the algebraic 2D invariants to 3D and explicitly derived the second order moment invariants, which were later reproduced by Guo \cite{Guo1993}. Using a group theoretic approach, Lo and Don \cite{LoDon1989} constructed twelve complex moment invariants including both second and third order moments. Galvez and Canton \cite{GalvezCanton1992} defined the 3D moments by evaluating them on the 3D object's surface and extracting global descriptors from normalized surface shapes. An extension of moment invariants to $n$-dimension can be found in Mamistvalov's work \cite{Mamistvalov1998}, in which the zeroth and second order moment invariants of $n$-dimensional regular solids were established. Other examples of 3D moments that are invariant to rotation and blur were provided by Flusser \textit{et al.} \cite{FlusserEtAl2003}.

We have recently developed a set of 2D local moment invariants based on the discrete Krawtchouk polynomials and successfully applied them to the comparison of local image patches \cite{SitKihara2014}. Krawtchouk polynomials were introduced by Mikhail Krawtchouk \cite{Pryzva1992} and used for the first time in image analysis by Yap \textit{et al.} \cite{YapRaveendranOng2002}. They have been employed in image processing and pattern recognition fields including image reconstruction \cite{YapParamesranOng2003}, image watermarking \cite{VenkataramanaRaj2007}, and face recognition \cite{JassimRaveendran2012,RaniDevaraj2012}. In our previous 2D work \cite{SitKihara2014}, while constructing a set of local descriptors that are rotation, position, and size independent, we have also preserved their ability to extract features from any local interest region in an image.

In this paper, we extend our previous 2D work to 3D for local comparison of 3D surface shapes. Our new method is based on 3D Krawtchouk polynomials. 3D Krawtchouk moments were earlier defined and used in content-based search and retrieval of 3D objects \cite{MademlisEtAl2006,DarasEtAl2006,XiangEtAl2006}. Despite the compact representation and discriminative powers of these moments, the theory of invariants based on 3D Krawtchouk polynomials has not been well studied. Also, the very critical local retrieval property of the 3D moments has been noticed in \cite{MesbahEtAl2016}, but much of the focus is given to their fast computation.
We propose a new approach on this long-standing issue of local image comparison by constructing 3D Krawtchouk descriptors (3DKD) for describing local 3D surfaces. The new formulation has many advantages over many similar moment-based approaches, such as TRS invariants \cite{FlusserEtAl2009} and Zernike descriptors \cite{NovotniKlein2003}: 1) Krawtchouk polynomials are defined on a discrete space, so the moments derived from them do not carry any error due to discretization unlike many other moments related to continuous functions \cite{FlusserEtAl2016}. 2) These polynomials are orthogonal; each moment brings in a new feature of the image, where minimum redundancy is critical in their discriminative performance. Moreover, they are directly defined in the image coordinate space, and hence their orthogonality property is well retained in the computed moments. 3) They are complete with a finite number of functions (equal to the image size) while many other polynomial spaces have infinitely many members. 4) They have the ability to retrieve local image patches by only changing the resolution of reconstruction and using low order moments. 5) The location of the patch can also be controlled by changing three parameters and hence shifting the region-of-interest along each dimension. 6) We also prove that these moments can be transformed into local descriptors, which are invariant under translation, rotation, and scaling. Therefore, using only a small number of invariant descriptors per image will make it possible to develop an efficient method for quick local image retrieval.

Moment-based approaches, particularly Krawtchouk moments, are very useful for representing biological and medical images as they are pixelized or voxelized data. In medical imaging, such as computerized tomography (CT) scan and magnetic resonance imaging (MRI), objects are observed at different viewpoints and local images need to be extracted and examined. In digital pathology, for instance, pathologists are interested in information about specific structures rather than the whole image \cite{MehtaEtAl2009}. Thus, it is necessary to construct moment invariants that do not change by translation, rotation, and scaling and can retrieve local image patches or subimages.

Local shape search methods have many applications also in structural biology, which deals with 3D structures of biomolecules. An important application is the identification of ligand molecules (i.e., small chemical compounds including drug molecules) that bind to local protein surface regions, which is important for predicting biological function of proteins \cite{ChikhiEtAl2011,SealKihara2012,ZhuEtAl2015} and for computational drug design \cite{RosenbergGoldblum2006,Patch,PLPS2}. Ligand molecules that bind to a local surface region in a protein can be predicted by finding similar local regions (binding pockets) of known ligand-binding proteins in the protein structure database.
In this work, we applied the developed 3DKD for the protein ligand binding pocket comparison. A ligand binding pocket is represented as a combination of overlapping local surface patches, each of which is characterized by its geometric shape. The shapes of surface patches are compactly represented by 3DKD. The method is benchmarked on a dataset, which contains a total of 463 proteins that bind to one of 11 ligand molecules. The 3DKD-based method shows higher prediction accuracy than a previously developed method, Pocket-Surfer \cite{ChikhiSaelKihara2010}, for seven of the 11 ligand types.

This paper is organized as follows. In Section \ref{sec:KrawPolys}, we give a brief background of one-dimensional Krawtchouk polynomials. Then, after introducing the 3D weighted Krawtchouk polynomials and their moments in Section \ref{sec:WeightedKrawPolys}, we present the theory and formulation of our new 3D Krawtchouk descriptors in Section \ref{sec:3DKD}. In Section \ref{sec:Computation}, we provide a detailed scheme for efficient computation of these descriptors. In Section \ref{sec:Results}, we show numerical results from local surface recognition performances of 3DKD using protein structures placed in different orientations. Finally, we discuss the application of 3DKD on the comparison of ligand binding pockets on protein surfaces. We finish the paper with a conclusion and summary of this work in Section \ref{sec:Conc}.

\section{Krawtchouk Polynomials}
\label{sec:KrawPolys}

We start with introducing one-dimensional Krawtchouk polynomials, which can also be found in \cite{YapRaveendranOng2002,YapParamesranOng2003,ZhangEtAl2010}. A more general and abstract form of these polynomials was provided as Hahn polynomials in \cite{YapParamesranOng2007}.

The $n$th order Krawtchouk polynomials are defined as
\begin{equation}
	\begin{split}
		K_n(x;p,N) & = \sum_{i=0}^{n} a_{i,n,p,N}  \, x^i  = {}_{2} F_{1} (-n,-x;-N;\frac{1}{p})
	\end{split}
	\label{eqn:KrawPolys}
\end{equation}
where $x,n=0,\ldots,N$, $N>0$, $p\in (0,1)$ and the function ${}_{2} F_{1}$ is the hypergeometric function which is defined as:
\begin{equation}
  {}_{2} F_{1} (a,b;c;z) = \sum_{i=0}^{\infty} \frac{(a)_i(b)_i}{(c)_i}\frac{z^i}{i!}.
  \label{eqn:HypergeometricFuncs}
\end{equation}
The symbol $(a)_i$ in (\ref{eqn:HypergeometricFuncs}) is the Pochhammer symbol given by
\begin{equation}
  (a)_i = a(a+1)(a+2)\ldots (a+i-1) = \frac{\Gamma(a+i)}{\Gamma(a)}.
  \label{eqn:PochhammerSymb}
\end{equation}

Note that the series in (\ref{eqn:HypergeometricFuncs}) terminates if either $a$ or $b$ is a nonpositive integer. Hence, the polynomial coefficients $a_{i,n,p,N}$ in (\ref{eqn:KrawPolys}) can be obtained by simplifying the summation. The first three Krawtchouk polynomials are
\begin{align*}
  	K_0(x;p,N)  & = 1, \\
	K_1(x;p,N)  & = 1 - \left(\frac{1}{Np}\right)x, \\
	K_2(x;p,N)  & = 1 - \left(\frac{2}{Np} + \frac{1}{N(N-1)p^2}\right)x \\
	            & \;\;\; + \left(\frac{1}{N(N-1)p^2}\right)x^2.
  \label{eqn:firstThreePolys}
\end{align*}

It is shown in \cite{YapParamesranOng2003} that the range of Krawtchouk polynomials expands rapidly with the increase of the order. Besides, these polynomials are not numerically stable for large values of $N$. Hence, a more stable set of polynomials can be obtained from the classical Krawtchouk polynomials by normalizing with the norm and scaling by the square root of a weight function \cite{YapParamesranOng2003}. The weighted Krawtchouk polynomials is then defined by
\begin{equation}
  \bar{K}_n(x;p,N) = {K}_n(x;p,N) \sqrt{ \frac{w(x;p,N)}{\rho(n;p,N)} }
  \label{eqn:WeightedKrawPolys}
\end{equation}
where
\begin{align}
  w(x;p,N)  & = {\binom{N}{x}} p^x (1-p)^{N-x}, \\
	\rho(n;p,N)  & = (-1)^n \left(\frac{1-p}{p}\right)^n \frac{n!}{(-N)_n}.
  \label{eqn:Rho}
\end{align}
The set of weighted Krawtchouk polynomials
\begin{equation}
  \bar{S} = \{ \bar{K}_n (x;p,N) : n = 0,\ldots,N \}
  \label{eqn:SetofWeightedKrawPolys}
\end{equation}
becomes a complete orthonormal set of basis functions on the discrete space $\{ 0,\ldots,N\}$ with the orthonormality condition
\begin{equation}
  \sum_{x=0}^{N} \bar{K}_n(x;p,N) \bar{K}_{n'}(x;p,N) = \delta_{nn'}.
  \label{eqn:Orthonormality}
\end{equation}

To compute the weighted Krawtchouk polynomials, the three-term recurrence relation given in \cite{YapParamesranOng2003} can be used. Such a recursive computation is shown to be more efficient than computing high order polynomials directly using (\ref{eqn:KrawPolys}) and (\ref{eqn:WeightedKrawPolys}). However, due to error propagation, computing polynomials recursively may still be numerically unstable for large $N$ as noted by Zhang \textit{et al.} \cite{ZhangEtAl2010}. To achieve numerical stability, we use symmetry and bi-recursive algorithm given in \cite{ZhangEtAl2010}.

\section{Three-dimensional Weighted Krawtchouk Moments}
\label{sec:WeightedKrawPolys}

In this section, we give a brief formulation of 3D weighted Krawtchouk moments, which are also introduced in \cite{MademlisEtAl2006,DarasEtAl2006,XiangEtAl2006}. Note that the functions $\bar{K}_n$ defined by (\ref{eqn:WeightedKrawPolys}) are orthonormal in the one-dimensional discrete set $\{0,\ldots,N\}$, but they can be easily extended to three-dimension as follows:

Let
\begin{equation}
  A = \{0,\ldots,N\} \times \{0,\ldots,M\} \times \{0,\ldots,L\}
  \label{eqn:3DDiscreteDomain}
\end{equation}
be a discrete field in the 3D space. We define the set of 3D weighted Krawtchouk polynomials on $A$ as
\begin{equation}
  \begin{split}
		\bar{S} = & \,\{ \bar{K}_n (x;p_x,N) \cdot \bar{K}_m (y;p_y,M) \cdot \bar{K}_l (z;p_z,L): \\
		          & \, n = 0,\ldots,N, \; m = 0,\ldots,M,  \; l = 0,\ldots,L \}.
		\end{split}
  \label{eqn:SetofWeighted3DKrawPolys}
\end{equation}

Note that $\bar{S}$ is orthonormal on $A$ with the orthonormality condition
\begin{equation}
	\begin{split}
		       \;\sum_{x=0}^{N} \sum_{y=0}^{M} & \sum_{z=0}^{L} \,\bar{K}_n (x;p_x,N) \, \bar{K}_m (y;p_y,M) \, \bar{K}_l (z;p_z,L) \\
	  \cdot & \;\bar{K}_{n'} (x;p_x,N) \, \bar{K}_{m'} (y;p_y,M) \, \bar{K}_{l'} (z;p_z,L) \\
	      =  &   \;\delta_{nn'} \delta_{mm'} \delta_{ll'},
	\end{split}
  \label{eqn:Orthonormality3D}
\end{equation}
which follows immediately from the orthonormality of 1D functions given by (\ref{eqn:Orthonormality}).

Let $f(x,y,z)$ be a 3D function defined on the grid $A$ in (\ref{eqn:3DDiscreteDomain}). The 3D weighted Krawtchouk moments of order $n+m+l$ of $f(x,y,z)$ are defined by
\begin{equation}
	\begin{split}
		\bar{Q}_{nml} = & \,\sum_{x=0}^{N} \sum_{y=0}^{M} \sum_{z=0}^{L} f(x,y,z) \\
		          \cdot & \,\bar{K}_n (x;p_x,N) \, \bar{K}_m (y;p_y,M) \, \bar{K}_l (z;p_z,L).
	\end{split}
  \label{eqn:Weighted3DKrawMoments}
\end{equation}

Note that by using (\ref{eqn:Orthonormality3D}) and solving (\ref{eqn:Weighted3DKrawMoments}) for $f(x,y,z)$, the 3D function $f(x,y,z)$ can be written in terms of the 3D weighted Krawtchouk polynomials, i.e.,
\begin{equation}
	\begin{split}
		f(x,y,z) = & \sum_{n=0}^{N} \sum_{m=0}^{M} \sum_{l=0}^{L} \,\bar{Q}_{nml} \\
		     \cdot & \, \bar{K}_n (x;p_x,N) \, \bar{K}_m (y;p_y,M) \, \bar{K}_l (z;p_z,L).
	\end{split}
  \label{eqn:fintermsof3DKrawMoments}
\end{equation}

This means that the object $f(x,y,z)$ can be reconstructed perfectly if all the moments $\bar{Q}_{nm}$ are used for $n=0,\ldots,N$, $m=0,\ldots,M$, $l=0,\ldots,L$. An approximate reconstruction $\hat{f}(x,y,z)$ of $f(x,y,z)$ can be written as
\begin{equation}
	\begin{split}
		\hat{f}(x,y,z) = & \sum_{n=0}^{\hat{N}} \sum_{m=0}^{\hat{M}} \sum_{l=0}^{\hat{L}} \,\bar{Q}_{nml} \\
		\cdot & \,\bar{K}_n (x;p_x,N) \, \bar{K}_m (y;p_y,M) \, \bar{K}_l (z;p_z,L).
	\end{split}
  \label{eqn:fhatintermsof3DKrawMoments}
\end{equation}
where $0\leq\hat{N}\leq N$, $0\leq\hat{M}\leq M$, $0\leq\hat{L}\leq L$.


\begin{figure*}[!ht]
		\centering
		\scriptsize
		\begin{tabular}{c|cccc}
        Voxelized 3D Image & $\hat{N}=\hat{M}=\hat{L}=5$ & $\hat{N}=\hat{M}=\hat{L}=10$ & $\hat{N}=\hat{M}=\hat{L}=25$ & $\hat{N}=\hat{M}=\hat{L}=50$\\
        \hline\\
				\hspace{-0.4cm}
       \begin{tikzpicture}
  			\node (img) {\includegraphics[scale=0.206,trim = 0.9cm 1.5cm 0.9cm 1.5cm,clip=true]{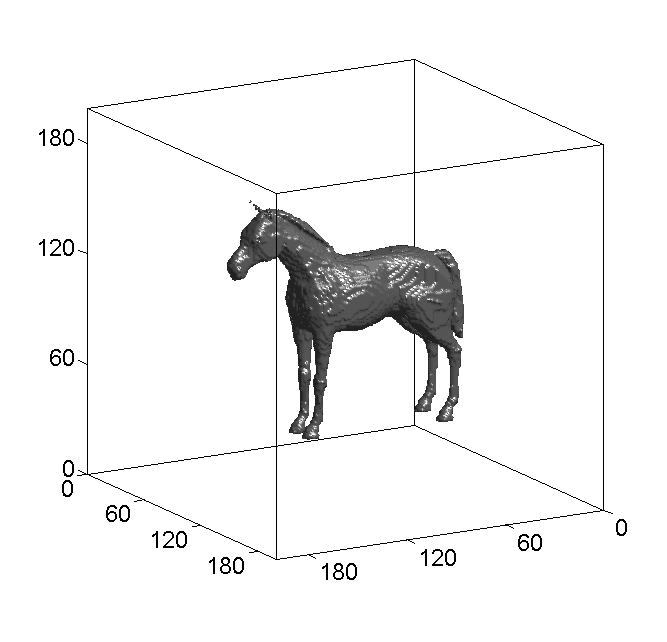}};
  			\node[below=of img, node distance=0cm, xshift=0.75cm , yshift=1.2cm,font=\color{black}] {$z$};
  			\node[below=of img, node distance=0cm, xshift=-1.05cm , yshift=1.1cm, anchor=center,font=\color{black}] {$x$};
  			\node[left=of img, node distance=0cm, xshift=0.9cm, yshift=0.2cm, anchor=center,font=\color{black}] {$y$};
		\end{tikzpicture}
         &
				\hspace{-0.2cm}
		\begin{tikzpicture}
        	\node (img) {\includegraphics[scale=0.206,trim = 0.9cm 1.5cm 0.9cm 1.5cm,clip=true]{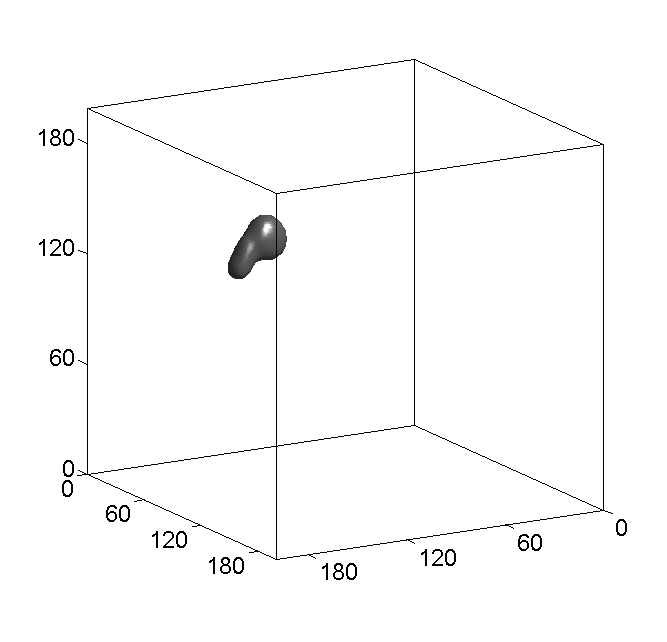}};
		\end{tikzpicture} 
		  &
			\hspace{-0.4cm}
		\begin{tikzpicture}
  			\node (img) {\includegraphics[scale=0.206,trim = 0.9cm 1.5cm 0.9cm 1.5cm,clip=true]{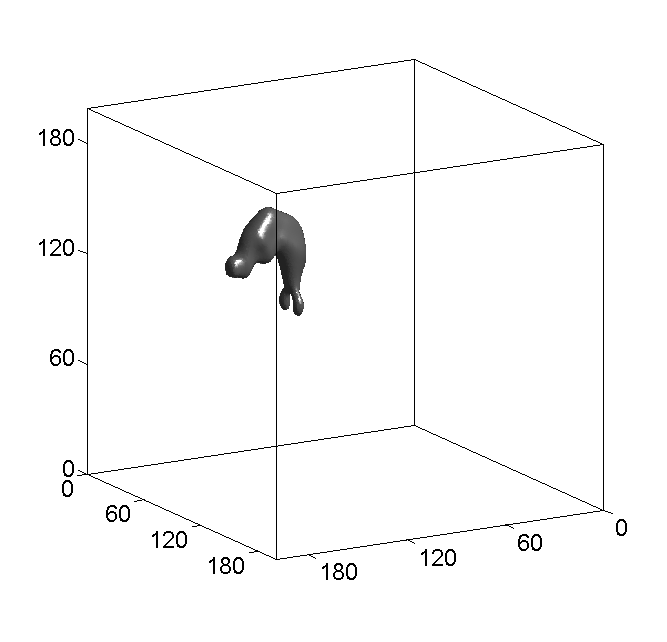}};
		\end{tikzpicture}
		  &
			\hspace{-0.4cm}
		\begin{tikzpicture}
  			\node (img) {\includegraphics[scale=0.206,trim = 0.9cm 1.5cm 0.9cm 1.5cm,clip=true]{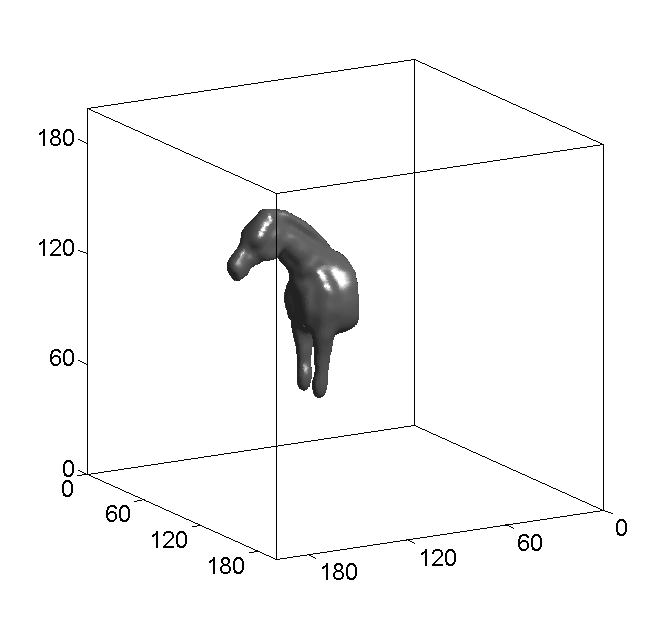}};
		\end{tikzpicture} 
		  &
			\hspace{-0.4cm}
		\begin{tikzpicture}
  			\node (img) {\includegraphics[scale=0.206,trim = 0.9cm 1.5cm 0.9cm 1.5cm,clip=true]{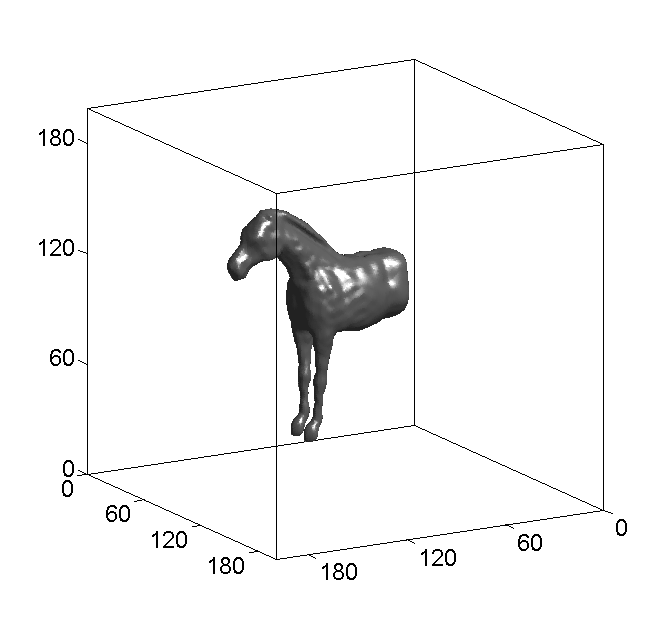}};
		\end{tikzpicture} 
				 \\
				\hspace{-0.4cm}
		  \begin{tikzpicture}
  			\node (img) {\includegraphics[scale=0.206,trim = 0.9cm 1.1cm 0.9cm 1.3cm,clip=true]{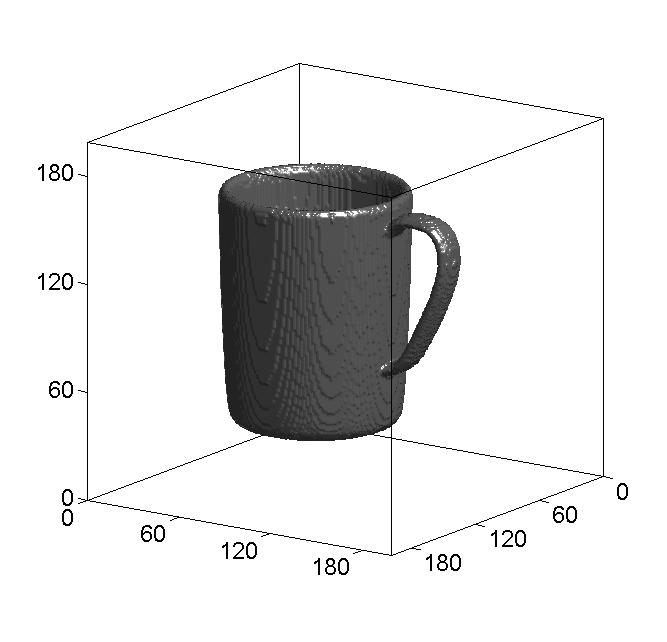}};
  			\node[below=of img, node distance=0cm, xshift=1.2cm , yshift=1.32cm,font=\color{black}] {$z$};
  			\node[below=of img, node distance=0cm, xshift=-0.7cm , yshift=1.1cm, anchor=center,font=\color{black}] {$x$};
  			\node[left=of img, node distance=0cm, xshift=0.9cm, yshift=0cm, anchor=center,font=\color{black}] {$y$};
		\end{tikzpicture} &
		\hspace{-0.2cm}
		\begin{tikzpicture}
  			\node (img) {\includegraphics[scale=0.206,trim = 0.9cm 1.1cm 0.9cm 1.3cm,clip=true]{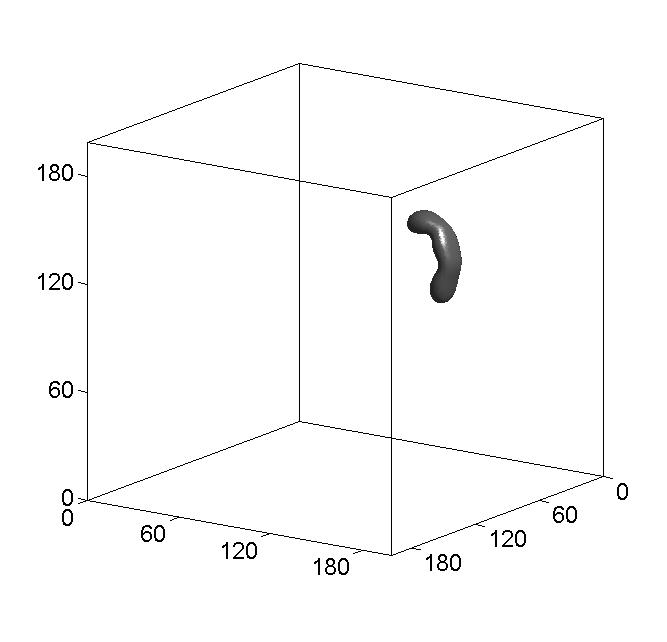}};
		\end{tikzpicture}  
		  &
			\hspace{-0.4cm}
		\begin{tikzpicture}
  			\node (img) {\includegraphics[scale=0.206,trim = 0.9cm 1.1cm 0.9cm 1.3cm,clip=true]{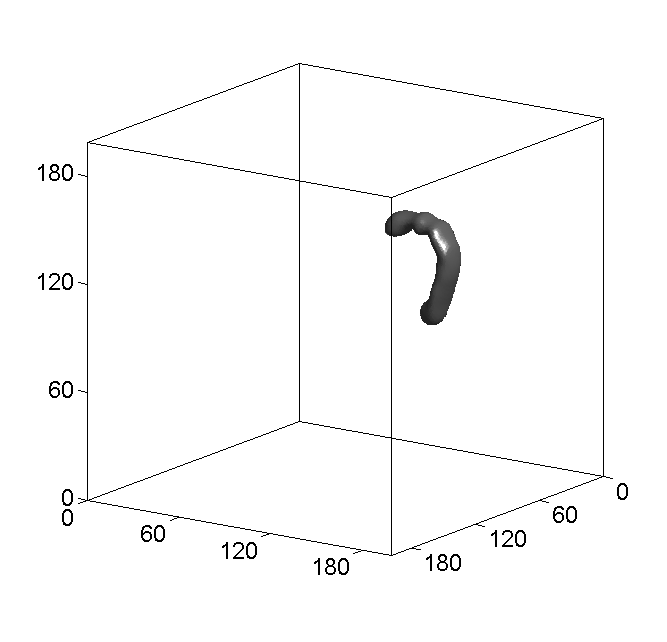}};
		\end{tikzpicture} 
		  &
			\hspace{-0.4cm}
		\begin{tikzpicture}
  			\node (img) {\includegraphics[scale=0.206,trim = 0.9cm 1.1cm 0.9cm 1.3cm,clip=true]{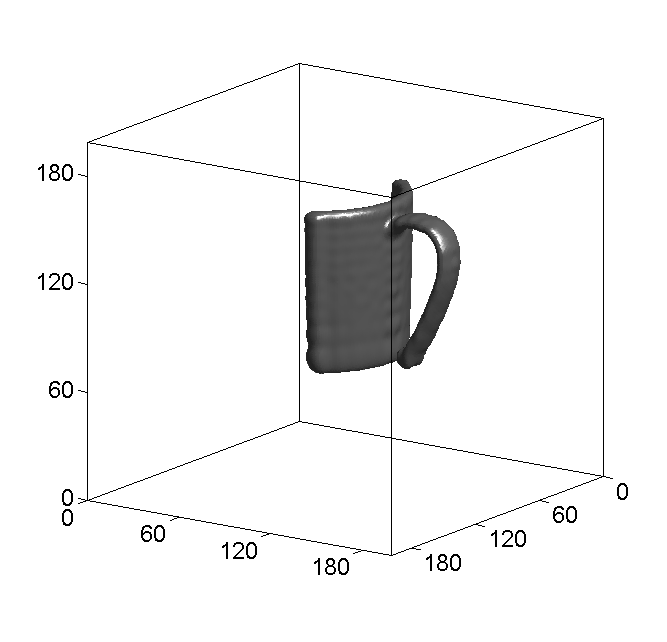}};
		\end{tikzpicture} 
		  &
			\hspace{-0.4cm}
		\begin{tikzpicture}
  			\node (img) {\includegraphics[scale=0.206,trim = 0.9cm 1.1cm 0.9cm 1.3cm,clip=true]{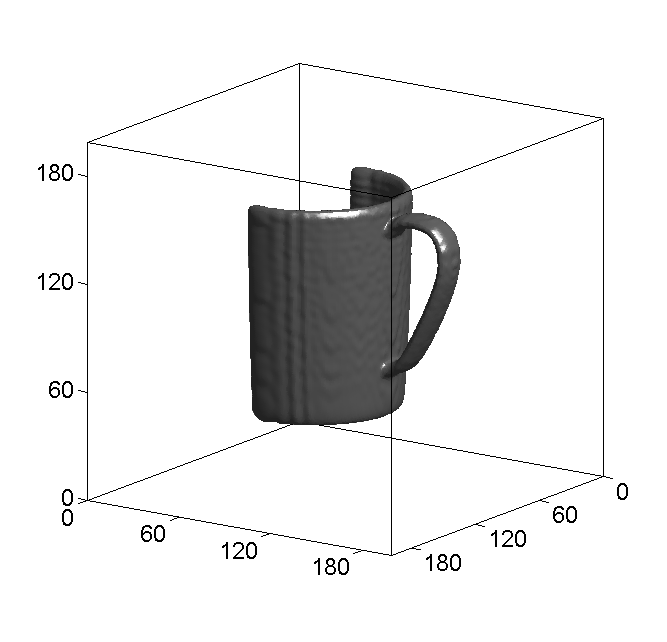}};
		\end{tikzpicture}
				 \\
				\hspace{-0.4cm}
        \begin{tikzpicture}
  			\node (img) {\includegraphics[scale=0.211,trim = 0.9cm 1cm 1.3cm 1.4cm,clip=true]{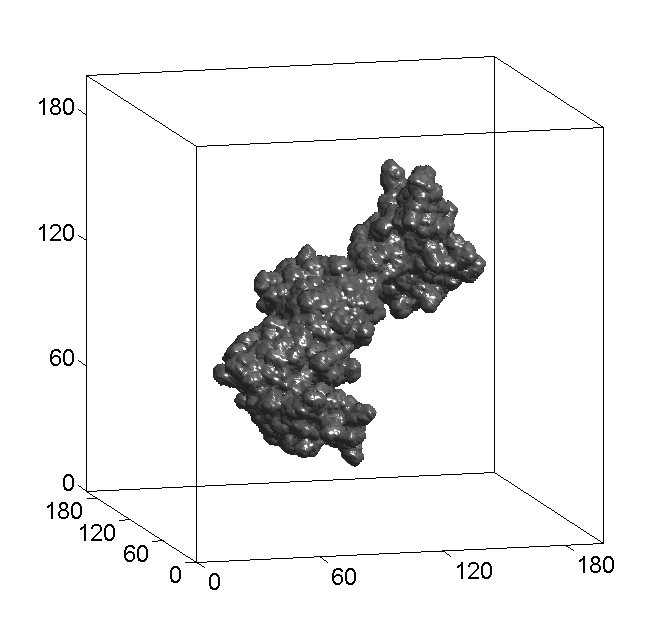}};
  			\node[below=of img, node distance=0cm, xshift=0.5cm , yshift=1.1cm,font=\color{black}] {$x$};
  			\node[below=of img, node distance=0cm, xshift=-1.3cm , yshift=1.15cm, anchor=center,font=\color{black}] {$y$};
  			\node[left=of img, node distance=0cm, xshift=0.9cm, yshift=0.2cm, anchor=center,font=\color{black}] {$z$};
		\end{tikzpicture} &
		\hspace{-0.2cm}
		\begin{tikzpicture}
  			\node (img) {\includegraphics[scale=0.211,trim = 0.9cm 1cm 1.3cm 1.4cm,clip=true]{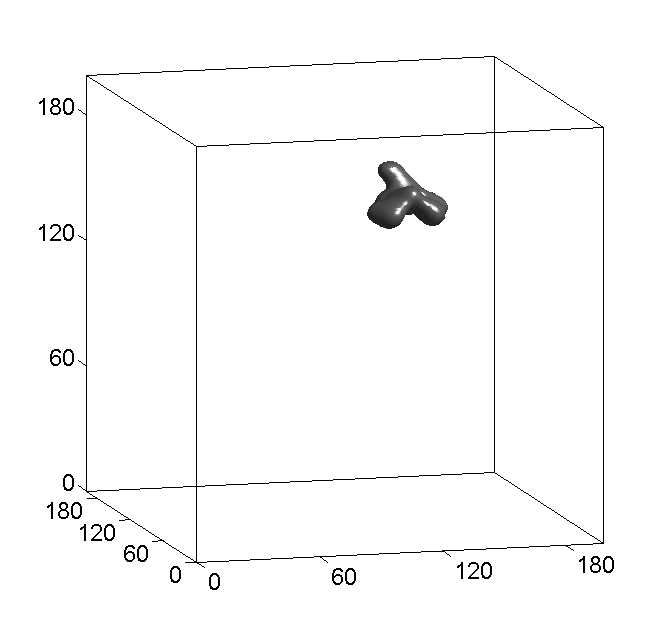}};
		\end{tikzpicture}  
		  &
			\hspace{-0.4cm}
		\begin{tikzpicture}
  			\node (img) {\includegraphics[scale=0.211,trim = 0.9cm 1cm 1.3cm 1.4cm,clip=true]{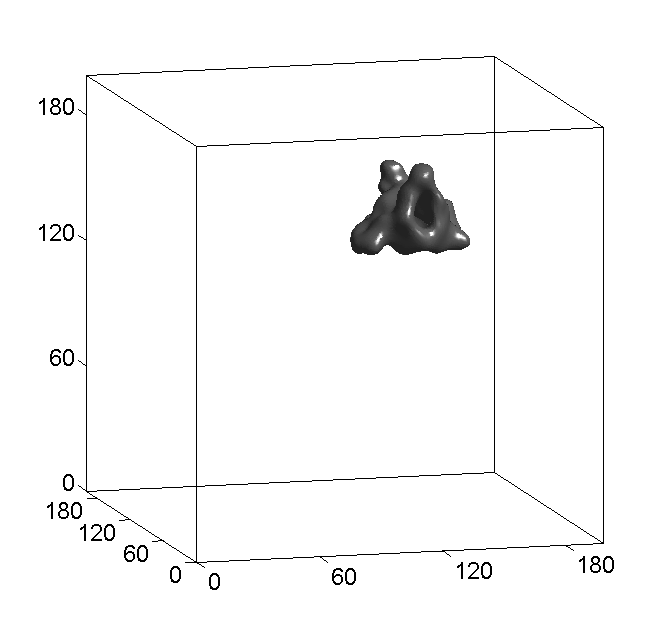}};
		\end{tikzpicture} 
		  &
			\hspace{-0.4cm}
		\begin{tikzpicture}
  			\node (img) {\includegraphics[scale=0.211,trim = 0.9cm 1cm 1.3cm 1.4cm,clip=true]{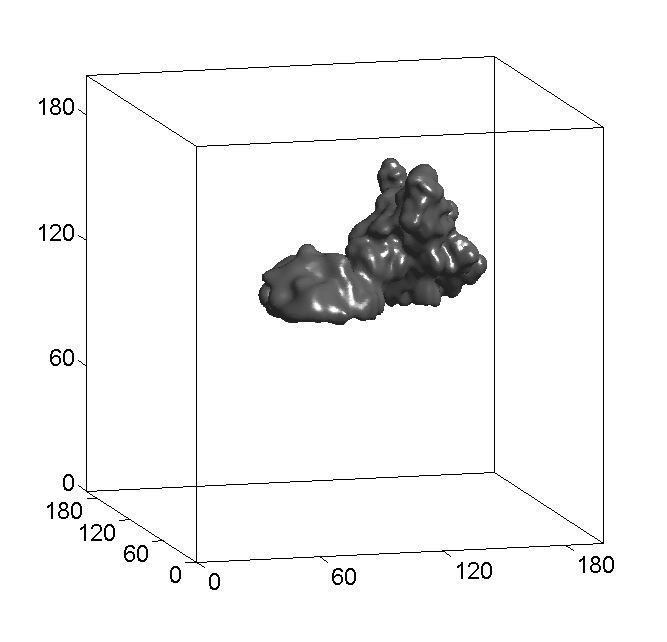}};
		\end{tikzpicture} 
		  &
			\hspace{-0.4cm}
		\begin{tikzpicture}
  			\node (img) {\includegraphics[scale=0.211,trim = 0.9cm 1cm 1.3cm 1.4cm,clip=true]{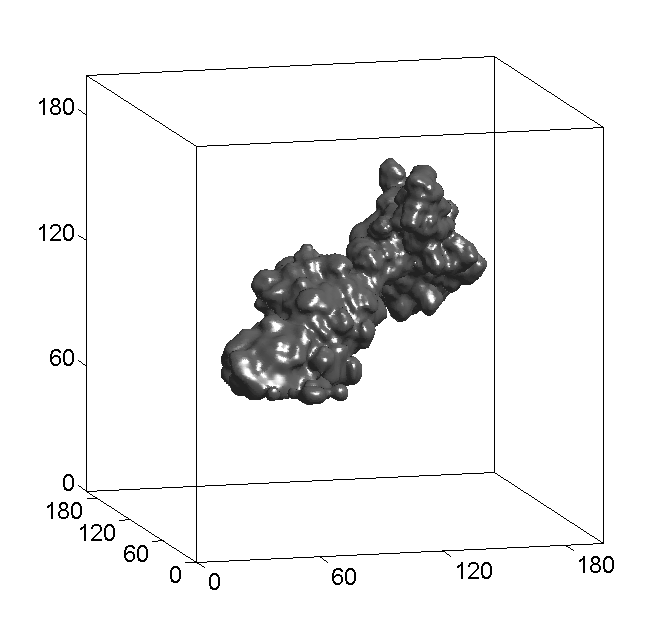}};
		\end{tikzpicture}\\
		\hspace{-0.4cm}
\begin{tikzpicture}
  			\node (img) {\includegraphics[scale=0.211,trim = 0.9cm 1cm 1.3cm 1.4cm,clip=true]{1ofc_voxel.png}};
  			\node[below=of img, node distance=0cm, xshift=0.5cm , yshift=1.1cm,font=\color{black}] {$x$};
  			\node[below=of img, node distance=0cm, xshift=-1.3cm , yshift=1.15cm, anchor=center,font=\color{black}] {$y$};
  			\node[left=of img, node distance=0cm, xshift=0.9cm, yshift=0.2cm, anchor=center,font=\color{black}] {$z$};
		\end{tikzpicture} &
		\hspace{-0.2cm}
		\begin{tikzpicture}
  			\node (img) {\includegraphics[scale=0.211,trim = 0.9cm 1cm 1.3cm 1.4cm,clip=true]{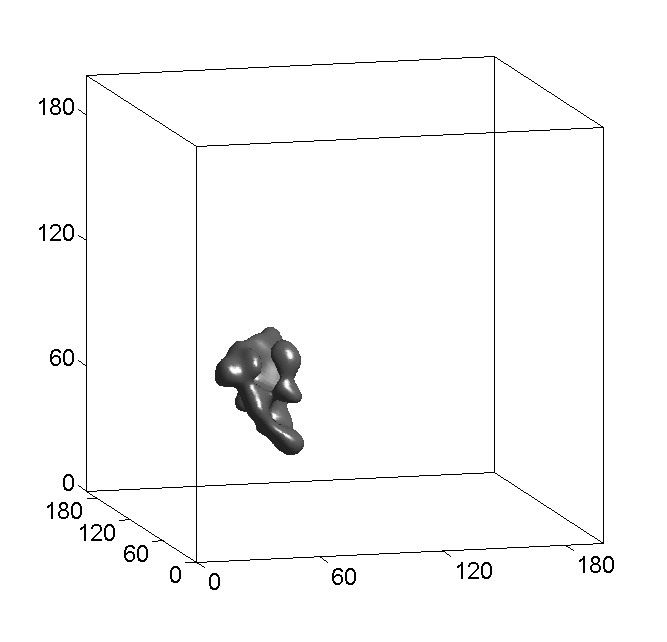}};
		\end{tikzpicture}  
		  &
			\hspace{-0.4cm}
		\begin{tikzpicture}
  			\node (img) {\includegraphics[scale=0.211,trim = 0.9cm 1cm 1.3cm 1.4cm,clip=true]{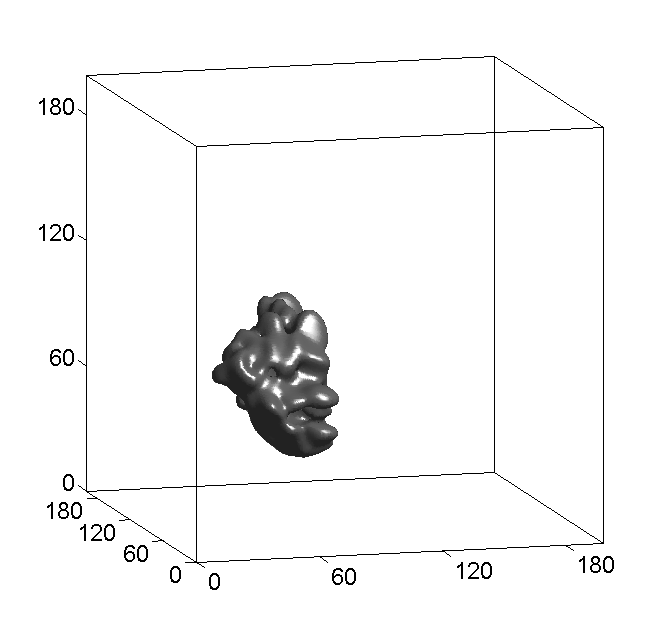}};
		\end{tikzpicture} 
		  &
			\hspace{-0.4cm}
		\begin{tikzpicture}
  			\node (img) {\includegraphics[scale=0.211,trim = 0.9cm 1cm 1.3cm 1.4cm,clip=true]{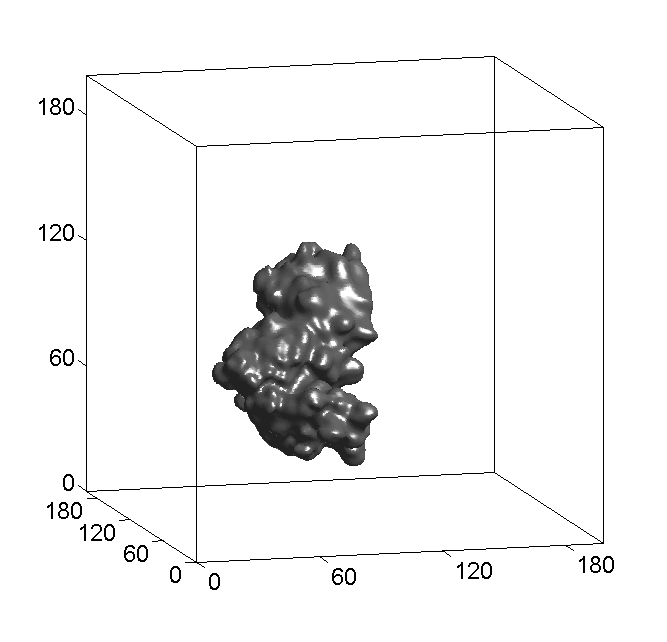}};
		\end{tikzpicture} 
		  &
			\hspace{-0.4cm}
		\begin{tikzpicture}
  			\node (img) {\includegraphics[scale=0.211,trim = 0.9cm 1cm 1.3cm 1.4cm,clip=true]{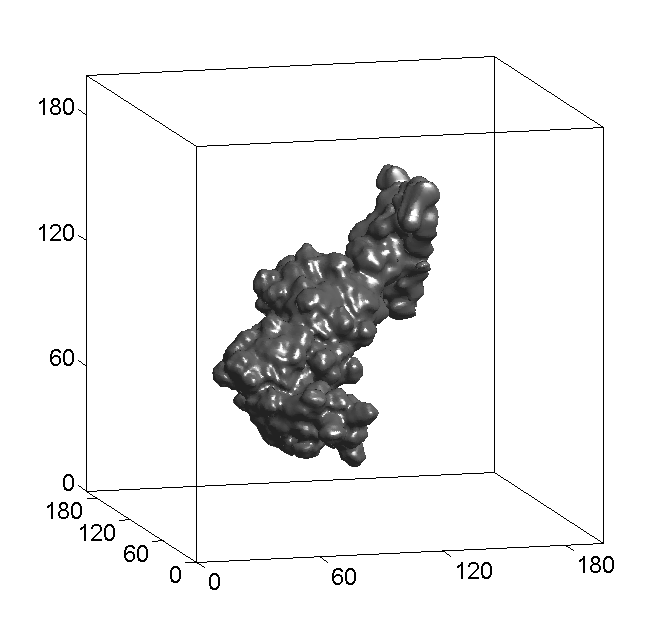}};
		\end{tikzpicture}\\
		\end{tabular}
    \caption{Examples of 3D binary images and their reconstructions using 3D weighted Krawtchouk polynomials for $\hat{N}=\hat{M}=\hat{L}=5$, $10$, $25$, and $50$, and different $(p_x,p_y,p_z)$ triplets. The voxel size for each box is $200^3$. $(p_x,p_y,p_z)$ triplet plays the critical role here in determining the center of local region-of-interest in an image. $(p_x, p_y, p_z)$ was set to (0.485, 0.630, 0.835) and (0.845, 0.725, 0.500) for the horse and the mug cup image, respectively, to obtain local images centered at the horse's mouth and the handle of the mug. The manually selected isosurface levels for the images from top to bottom are 0.33, 0.355, 0.195, and 0.195, respectively.}
		\label{fig:3Dexamples}
\end{figure*}









Fig.~\ref{fig:3Dexamples} presents some reconstructions of 3D binary images using 3D weighted Krawtchouk polynomials for $\hat{N}$, $\hat{M}$, $\hat{L}$ values of $5$, $10$, $25$, and $50$, and different $(p_x,p_y,p_z)$ triplets. The 3D polygonal models for the horse and the mug image are downloaded from Princeton Shape Benchmark \cite{PrincetonShapeBenchmark2004} and voxelized using the algorithm in \cite{PatilRavi2005}. The 3D weighted Krawtchouk moments from the original image are first computed using (\ref{eqn:Weighted3DKrawMoments}), and then these moments are used in (\ref{eqn:fhatintermsof3DKrawMoments}) for reconstructing the image.The center of a local region corresponding to $(p_x,p_y,p_z)$ is at $(x_c,y_c,z_c)$ where $x_c=N p_x$, $y_c=M p_y$, and $z_c=L p_z$. These points are $(97,126,167)$ near the horse's mouth and $(169,145,100)$ near the mug's handle. Since $N=M=L=200$ in this example, the $(p_x,p_y,p_z)$ triplets at these centers will correspond to $(0.485,0.630,0.835)$ and $(0.845,0.725,0.500)$, respectively.

As can be seen from left to right in Fig.~\ref{fig:3Dexamples}, the reconstructions start at a local region corresponding to $(Np_x, Mp_y, Lp_z)$ and expand as larger values of $\hat{N}$, $\hat{M}$, and $\hat{L}$ are used.
Theoretically, using $\hat{N}=N=200$, $\hat{M}=M=200$, and $\hat{L}=L=200$, the original image will be fully reconstructed regardless of the choice of $(p_x,p_y,p_z)$. Using smaller numbers for $\hat{N}$, $\hat{M}$, and $\hat{L}$, the reconstructed surfaces contain only local information which may actually be more useful for local comparison of 3D images. The parameters $p_x$, $p_y$, and $p_z$ play a vital role here to determine the center of local region-of-interest. 

In the third and fourth row of Fig.~\ref{fig:3Dexamples}, we show the voxelized surface of a protein, nucleosome recognition module of imitation SWI ATPase from fruit fly (\textit{Drosophila melanogaster}). The atomic structure of this protein is downloaded from the Protein Data Bank (PDB) \cite{PDB2000} (PDB ID: 1OFC) and then is voxelized using 3D-Surfer \cite{3DSURFER}. The mean radius\footnote{The mean radius is calculated as the average of all distances between the center of mass of the protein and each amino acid.} of this protein is about 27.8~\AA. In the grid shown, $1$ unit corresponds to 0.6~\AA. For this example, two $(p_x,p_y,p_z)$ triplets are selected: $(0.66,0.67,0.82)$ and $(0.21,0.44,0.31)$ in the third and the fourth row, respectively. So, the reconstruction centers will be $(x_c,y_c,z_c)=(132,134,164)$ and $(42,88,62)$, respectively, both chosen from salient parts on the surface of the protein. Again, using smaller numbers for $\hat{N}$, $\hat{M}$, and $\hat{L}$, the reconstructed surfaces will contain only local information. Local retrieval of structures may reveal important information about the function of a protein, and this may be used to locally compare protein structures in a large database and quickly identify their ligand-binding sites. This may be very useful for identifying biological functions of proteins and further computational drug design for target proteins. In this paper, we will employ $p_x$, $p_y$, and $p_z$ parameters for detecting the local region-of-interest by changing them between $0$ to $1$.


\begin{figure*}[!ht]
  \scriptsize
  \centering
    \begin{tabular}{cccccc}
      & \hspace{-0.3cm}
      \includegraphics[scale=0.216,trim = 1.6cm 14.3cm 1.6cm 1cm,clip=true]{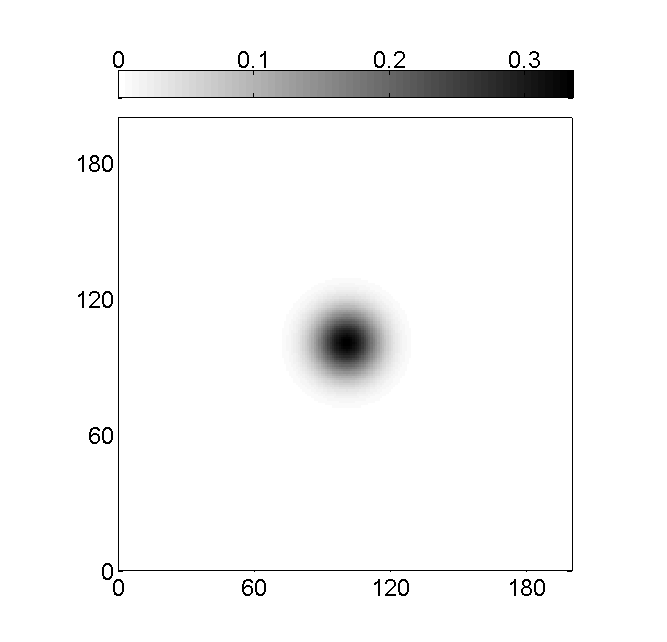} & 
      \includegraphics[scale=0.216,trim = 1.6cm 14.3cm 1.6cm 1cm,clip=true]{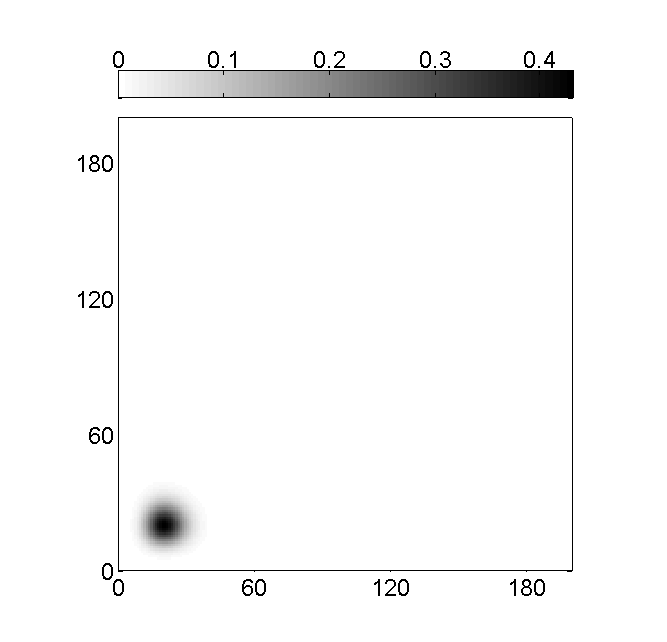} & 
      \includegraphics[scale=0.216,trim = 1.6cm 14.3cm 1.6cm 1cm,clip=true]{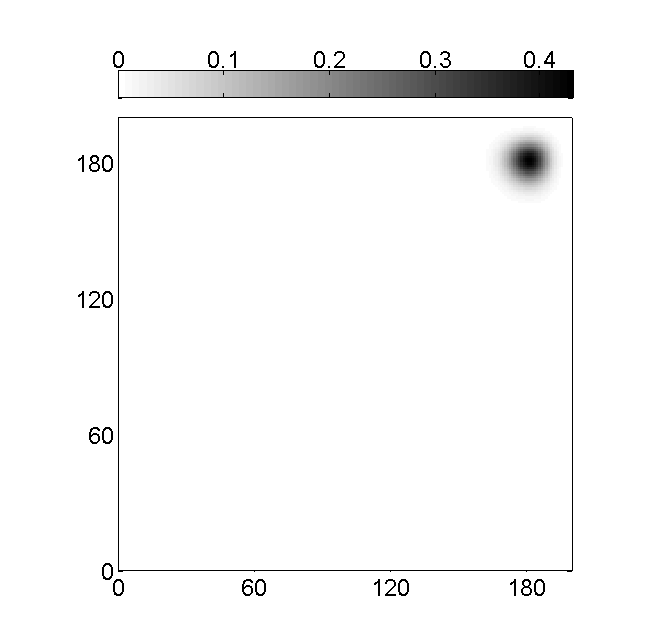} & 
      \includegraphics[scale=0.216,trim = 1.6cm 14.3cm 1.6cm 1cm,clip=true]{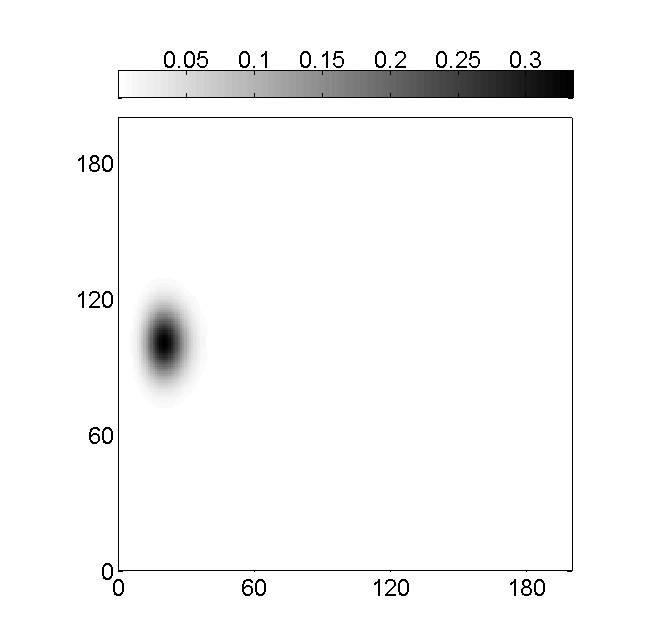} & 
      \includegraphics[scale=0.216,trim = 1.6cm 14.3cm 1.6cm 1cm,clip=true]{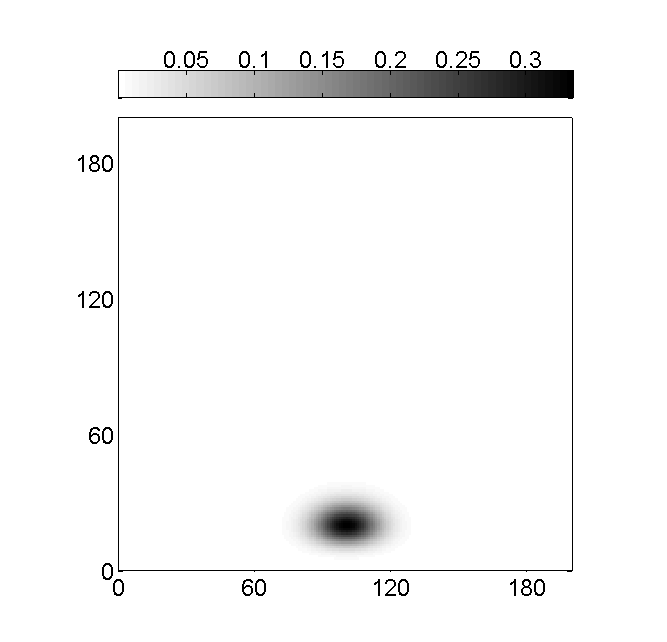}
      \\
      & $\mathbf{p} = (0.5, 0.5, 0.5)$ 
      & $\mathbf{p} = (0.1, 0.1, 0.1)$ 
      & $\mathbf{p} = (0.9, 0.9, 0.9)$  
      & $\mathbf{p} = (0.1, 0.5, 0.1)$
      & $\mathbf{p} = (0.5, 0.1, 0.9)$
      \\
      \rotatebox{90}{\hspace{1cm}$xy-$view} &\hspace{-0.3cm}
      \includegraphics[scale=0.2,trim = 1.2cm 1cm 1.2cm 1cm,clip=true]{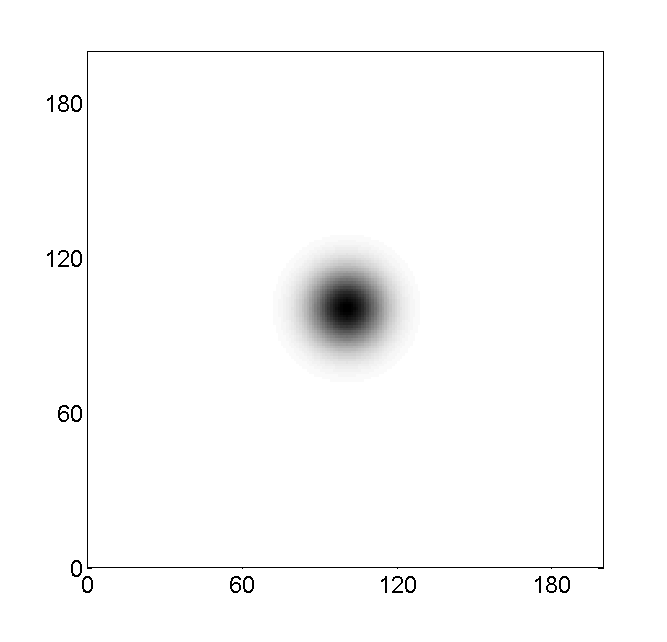} & 
      \includegraphics[scale=0.2,trim = 1.2cm 1cm 1.2cm 1cm,clip=true]{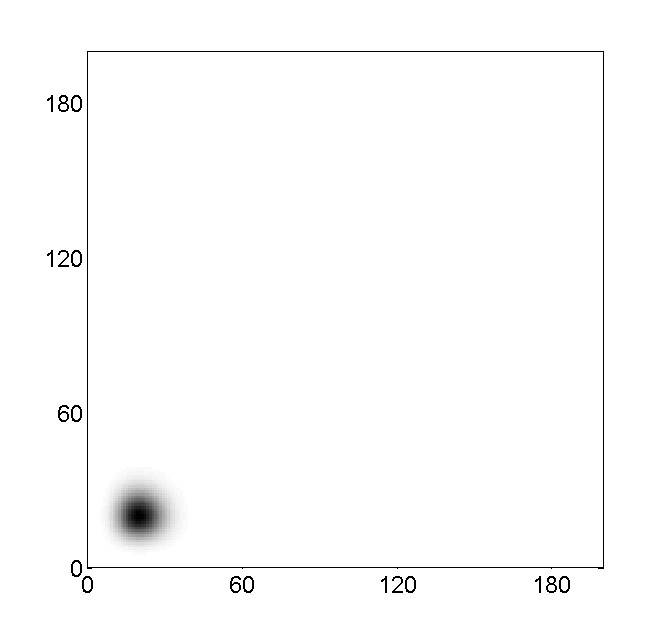} & 
      \includegraphics[scale=0.2,trim = 1.2cm 1cm 1.2cm 1cm,clip=true]{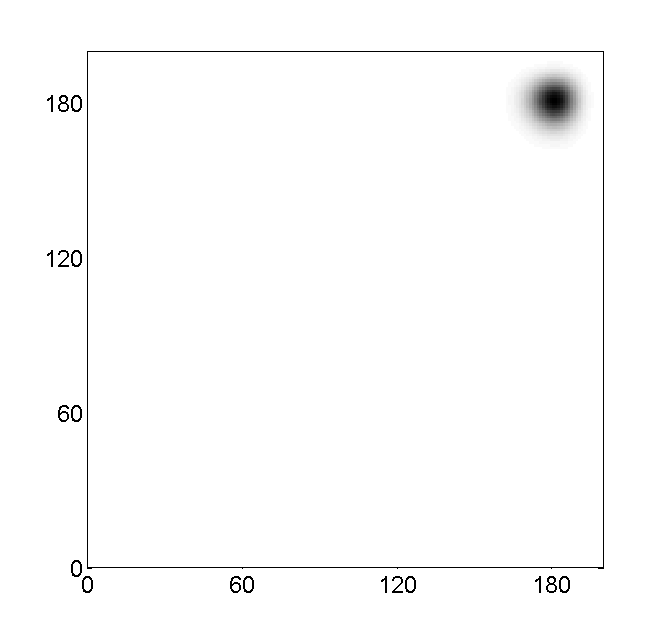} & 
      \includegraphics[scale=0.2,trim = 1.2cm 1cm 1.2cm 1cm,clip=true]{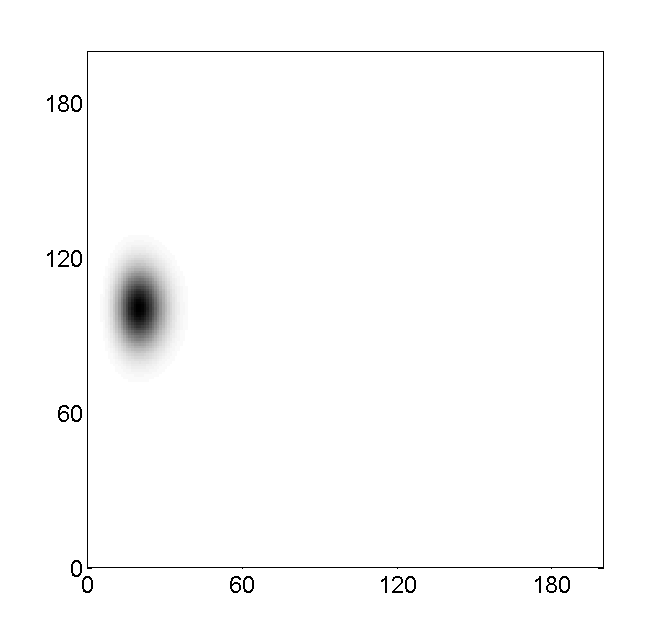} & 
      \includegraphics[scale=0.2,trim = 1.2cm 1cm 1.2cm 1cm,clip=true]{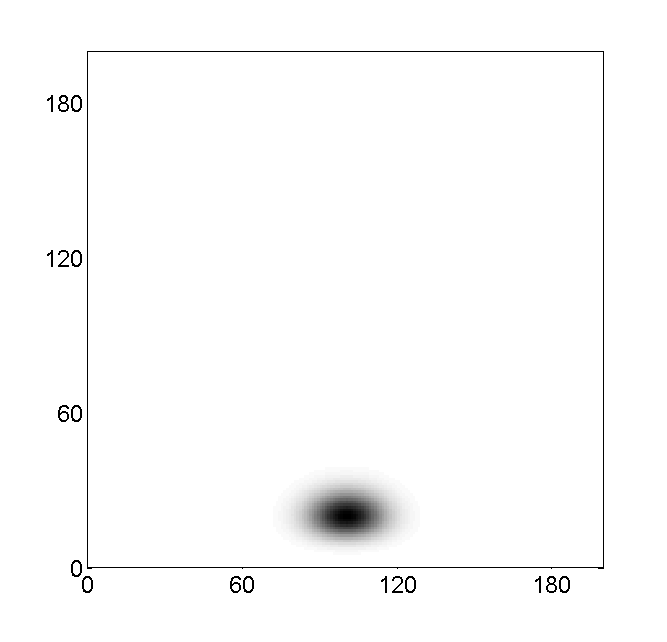}
      \\
      \rotatebox{90}{\hspace{1cm}$xz-$view} &\hspace{-0.3cm}
      \includegraphics[scale=0.2,trim = 1.2cm 1cm 1.2cm 1cm,clip=true]{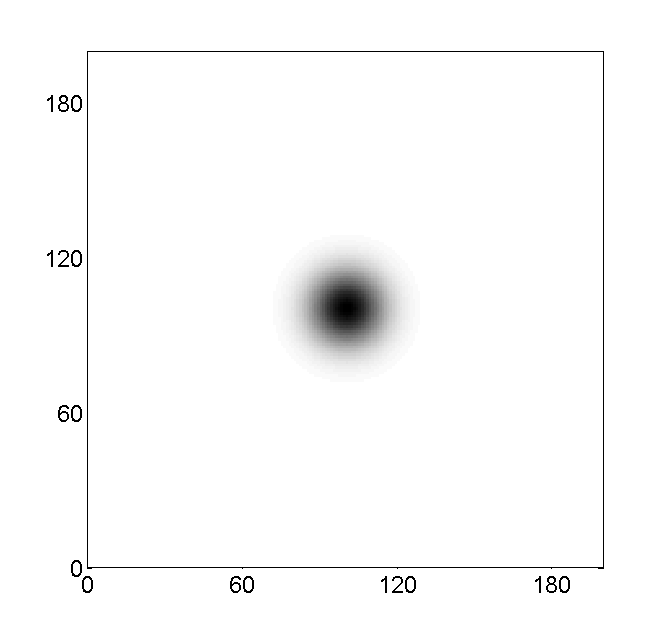} & 
      \includegraphics[scale=0.2,trim = 1.2cm 1cm 1.2cm 1cm,clip=true]{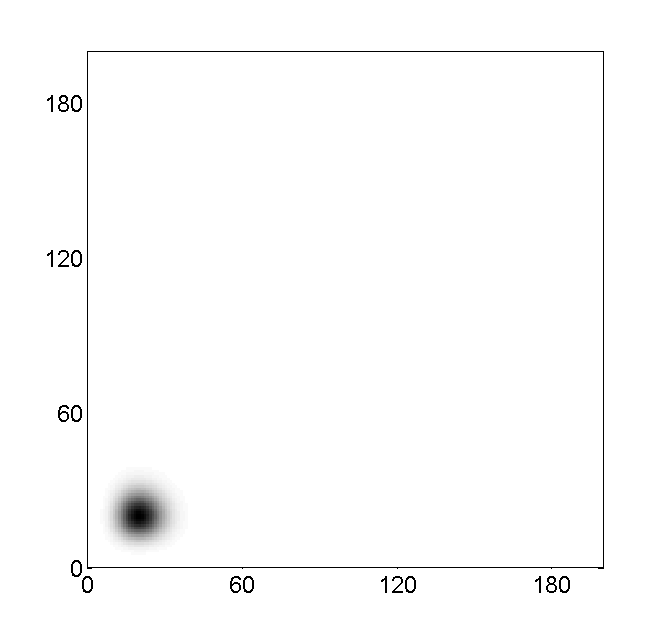} & 
      \includegraphics[scale=0.2,trim = 1.2cm 1cm 1.2cm 1cm,clip=true]{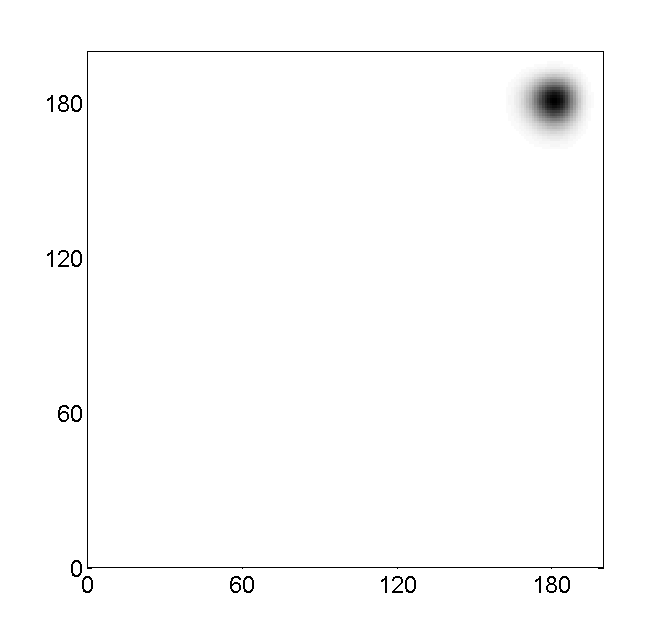} & 
      \includegraphics[scale=0.2,trim = 1.2cm 1cm 1.2cm 1cm,clip=true]{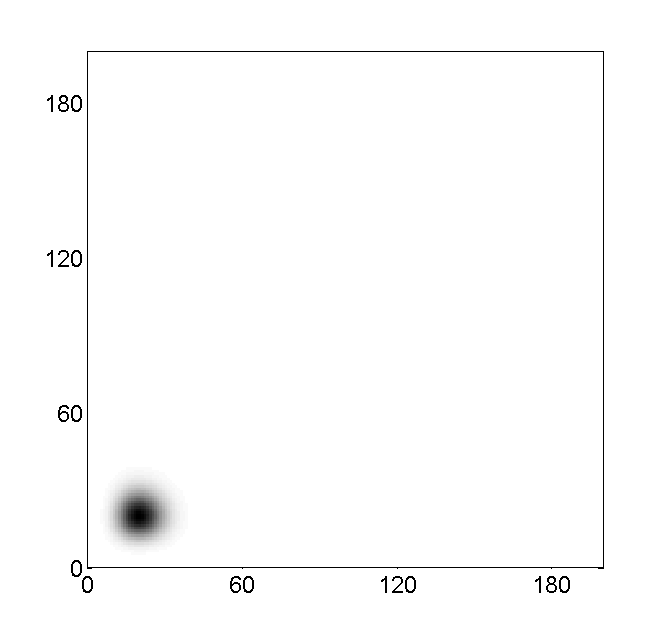} & 
      \includegraphics[scale=0.2,trim = 1.2cm 1cm 1.2cm 1cm,clip=true]{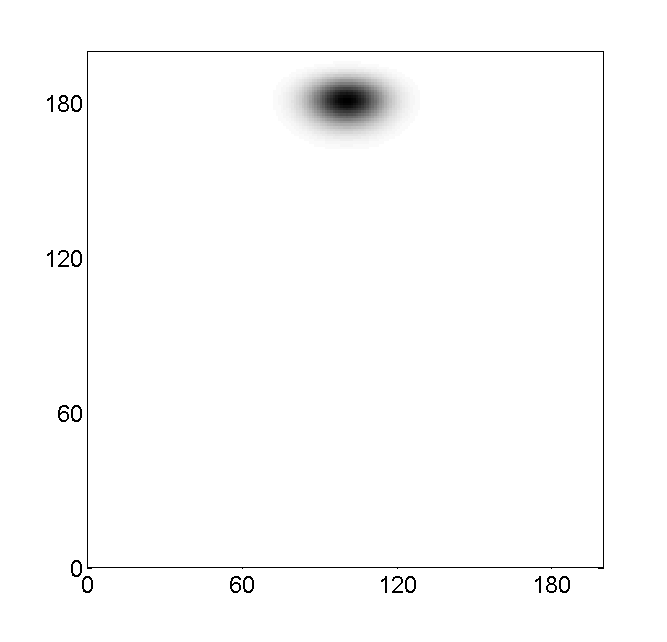}
      \\
      \rotatebox{90}{\hspace{1cm}$yz-$view} &\hspace{-0.3cm}
      \includegraphics[scale=0.2,trim = 1.2cm 1cm 1.2cm 1cm,clip=true]{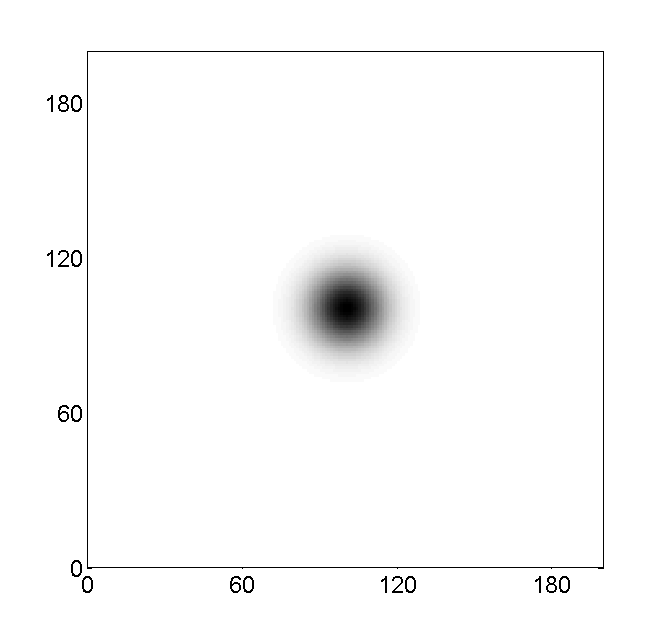} & 
      \includegraphics[scale=0.2,trim = 1.2cm 1cm 1.2cm 1cm,clip=true]{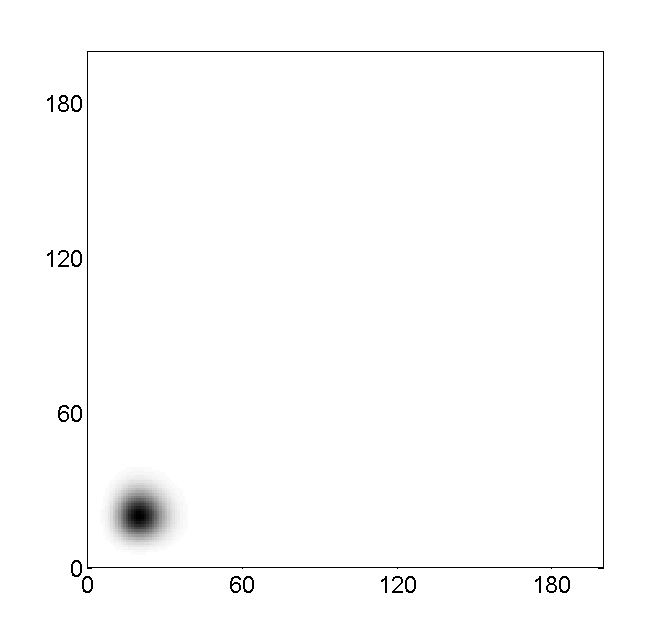} & 
      \includegraphics[scale=0.2,trim = 1.2cm 1cm 1.2cm 1cm,clip=true]{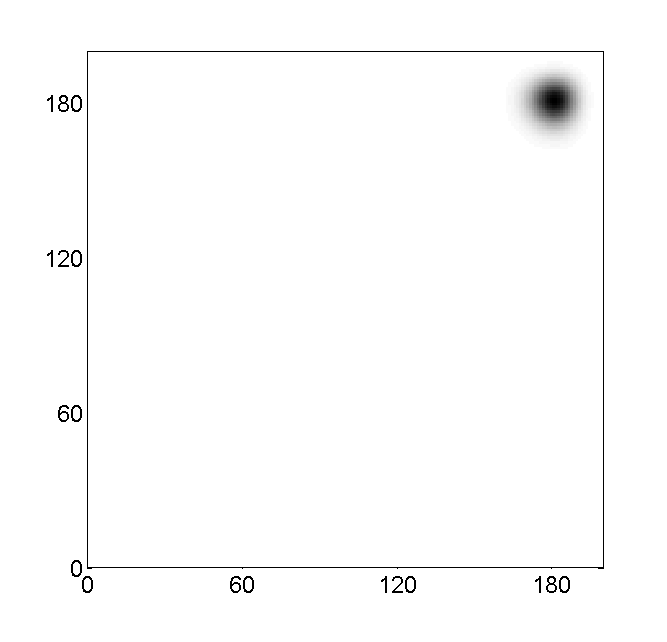} & 
      \includegraphics[scale=0.2,trim = 1.2cm 1cm 1.2cm 1cm,clip=true]{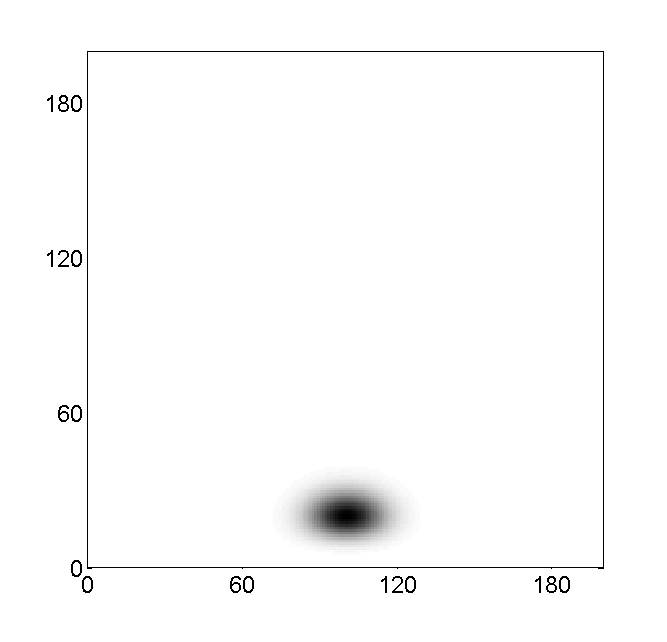} & 
      \includegraphics[scale=0.2,trim = 1.2cm 1cm 1.2cm 1cm,clip=true]{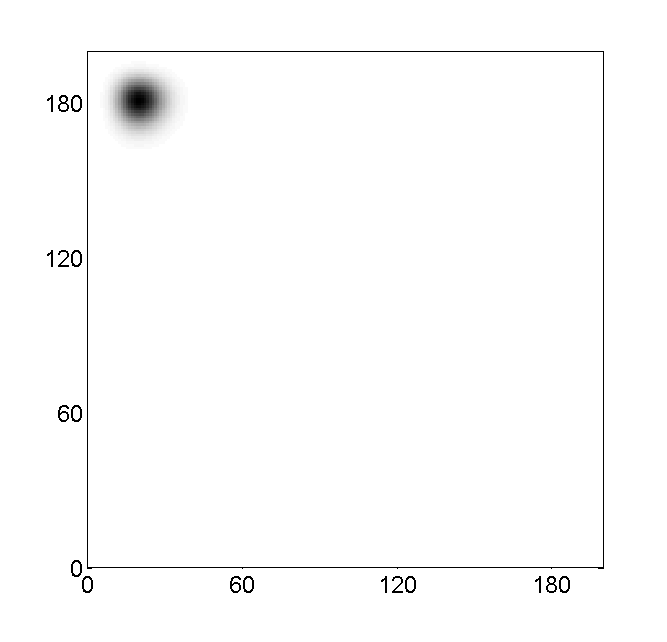}
      \\
    \end{tabular}
    \caption{\footnotesize{2D views of the square root of the weight function in (\ref{eqn:WeightedFunction}), flattened by summing slices along each dimension. The summations are performed along $z$, $y$, and $x$-axis, respectively from top to bottom. Density plots are shown for five different choices of $\mathbf{p}=(p_x,p_y,p_z)$ triplets. $\mathcal{N}=(200,200,200)$ in all cases. The gray scale colormaps at the top show the intensity of plots in each column.}}
    \label{fig:weights}
\end{figure*}

\section{3D Krawtchouk Descriptors}
\label{sec:3DKD}

In this section, we introduce a new set of invariants, called {\em 3D Krawtchouk descriptors}. We show that these invariants are not only rotation, size, and position independent, but also contain discriminative local features from any region-of-interest in a 3D image. Such invariants in 2D have been introduced in our previous work \cite{SitKihara2014}. In this work, we extend them to 3D to locally compare 3D images.

Let $f(x,y,z)$ be a function representing a 3D image defined on an orthogonal grid $A$ given in (\ref{eqn:3DDiscreteDomain}) and define the 3D weight function corresponding to triplets $\mathbf{p}=(p_x,p_y,p_z)$ and $\mathcal{N}=(N,M,L)$ by
\begin{equation}
	W(x,y,z;\mathbf{p},\mathcal{N}) = w(x;p_x,N)\,w(y;p_y,M)\,w(z;p_z,L).
	\label{eqn:3DWeightFunction}
\end{equation}
where $x=0,\ldots,N$, $y=0,\ldots,M$, and $z=0,\ldots,L$. Similarly, one can define the 3D norms corresponding to triplets $\mathbf{p}$ and $\mathcal{N}$ by
\begin{equation}
	\Omega (n,m,l;\mathbf{p},\mathcal{N})=\rho(n;p_x,N)\,\rho(m;p_y,M)\,\rho(l;p_z,L)
	\label{eqn:3DNorm}
\end{equation}
for $n=0,\ldots,N$, $m=0,\ldots,M$, and $l=0,\ldots,L$. Using (\ref{eqn:WeightedKrawPolys}), the 3D weighted Krawtchouk moments $\bar{Q}_{nml}$ in (\ref{eqn:Weighted3DKrawMoments}) become
\begin{equation}
	\begin{split}
		\bar{Q}_{nml} = & \, [\Omega (n,m,l;\mathbf{p},\mathcal{N})]^{-1/2} \cdot  \sum_{x=0}^{N} \sum_{y=0}^{M} \sum_{z=0}^{L} \,\tilde{f}(x,y,z)\\
						 & \cdot \, K_n (x;p_x,N) \, K_m (y;p_y,M) \, K_l (z;p_z,L)
	\end{split}
    \label{eqn:Weighted3DKrawMomentsv2}
\end{equation}
where
\begin{equation}
	\tilde{f}(x,y,z) = [W(x,y,z;\mathbf{p},\mathcal{N})]^{1/2}\, f(x,y,z).
	\label{eqn:WeightedFunction}
\end{equation} 
Now, substituting $K_n$, $K_m$, and $K_l$ in (\ref{eqn:Weighted3DKrawMomentsv2}) by their definitions from (\ref{eqn:KrawPolys}), reordering summations, and grouping terms, we obtain
\begin{equation}
  \begin{split}
   &\bar{Q}_{nml} =  \; [\Omega (n,m,l;\mathbf{p},\mathcal{N})]^{-1/2} \\
						\cdot & \sum_{i=0}^{n} \sum_{j=0}^{m} \sum_{k=0}^{l} \, a_{i,n,p_x,N} \, a_{j,m,p_y,M} \, a_{k,l,p_z,L} \, \tilde{\mathrm{M}}_{ijk},
  \end{split}
	\label{eqn:Qbar}
\end{equation}
where
\begin{equation}
  \tilde{\mathrm{M}}_{ijk} = \sum_{x=0}^{N} \sum_{y=0}^{M} \sum_{z=0}^{L} x^i y^j z^k \tilde{f}(x,y,z)
  \label{eqn:GeometricMomentsWeighted}
\end{equation}
are the geometric moments of the auxiliary function in (\ref{eqn:WeightedFunction}).

Notice that the geometric moments $\tilde{\mathrm{M}}_{ijk}$ and hence the weighted Krawtchouk moments $\bar{Q}_{nml}$ are not invariant under translation, rotation, and scaling. The translation invariant {\em central moments} of $\tilde{f}(x,y,z)$ can be defined as
\begin{equation}
	\begin{split}
  \tilde{\mu}_{nml} = \,\sum_{x=0}^{N} \sum_{y=0}^{M} \sum_{z=0}^{L} &\,\tilde{f}(x,y,z)\,(x-\tilde{x})^n (y-\tilde{y})^m (z-\tilde{z})^l ,
  \end{split}
  \label{eqn:tildeCentralMoments}
\end{equation}
where $\tilde{x}={\tilde{\mathrm{M}}_{100}}/{\tilde{\mathrm{M}}_{000}}$, $\tilde{y}={\tilde{\mathrm{M}}_{010}}/{\tilde{\mathrm{M}}_{000}}$, and $\tilde{z}={\tilde{\mathrm{M}}_{001}}/{\tilde{\mathrm{M}}_{000}}$ are the centroids of the auxiliary image $\tilde{f}(x,y,z)$. 

If $\tilde{\mu}_{nml}$ are the central moments, then  we can define geometric moments of $\tilde{f}(x,y,z)$, which are invariant under translation and scaling as follows:
\begin{equation}
	\tilde{\eta}_{nml} = \frac{\tilde{\mu}_{nml}}{(\tilde{\mathrm{M}}_{000})^{\frac{n+m+l}{3}+1}}.
	\label{eqn:tildeScaleInvMoments}
\end{equation}

Obtaining rotation invariant geometric moments of $\tilde{f}(x,y,z)$ is, however, not as straightforward. To achieve rotational invariance, we rotate the auxiliary image $\tilde{f}(x,y,z)$ so that its principal axes lie in the $x$, $y$, $z-$directions, respectively. When the auxiliary image $\tilde{f}(x,y,z)$ is centered at the origin, the principal axes of $\tilde{f}(x,y,z)$ can be defined as the eigenvectors of the inertia matrix 

\begin{equation}
\tilde{I} = \left[ 
\begin{array}{ccc}
   \;\,\,\tilde{I}_{xx} &  -\tilde{I}_{xy} & -\tilde{I}_{xz} \\
   -\tilde{I}_{yx} &  \;\,\,\tilde{I}_{yy} & -\tilde{I}_{yz} \\
   -\tilde{I}_{zx} &  -\tilde{I}_{zy} & \;\,\,\tilde{I}_{zz}
\end{array}
\right]
\end{equation}
where

\begin{equation}
  \begin{split}
    \tilde{I}_{xx} & = \tilde{\mu}_{020}+\tilde{\mu}_{002} \\
    \tilde{I}_{yy} & = \tilde{\mu}_{200}+\tilde{\mu}_{002} \\
    \tilde{I}_{zz} & = \tilde{\mu}_{200}+\tilde{\mu}_{020}
  \end{split}
\end{equation}
and

\begin{equation}
  \begin{split}
    \tilde{I}_{xy} & = \tilde{I}_{yx} = \tilde{\mu}_{110} \\
    \tilde{I}_{xz} & = \tilde{I}_{zx} = \tilde{\mu}_{101} \\
    \tilde{I}_{yz} & = \tilde{I}_{zy} = \tilde{\mu}_{011}.
  \end{split}
\end{equation}

Here, $\tilde{I}$ is a symmetric matrix with real eigenvalues $\{\tilde{\lambda}_1, \tilde{\lambda}_2, \tilde{\lambda}_3\}$ and orthogonal eigenvectors $\{\tilde{u}_1, \tilde{u}_2, \tilde{u}_3\}$ such that
\begin{equation}
  \begin{split}
    \tilde{I}\,\tilde{u}_i & = \tilde{\lambda}_i\,\tilde{u}_i \;\;\mbox{ for } \;\;i=1,2,3. \\
  \end{split}
\end{equation}
The eigenvectors $\{\tilde{u}_1, \tilde{u}_2, \tilde{u}_3\}$ define the rows of the rotation matrix $\tilde{R}$ that aligns the principal axes with the standard $xyz$ coordinate system. However, for each eigenvalue $\tilde{\lambda}_i$, both $\tilde{u}_i$ and $-\tilde{u}_i$ are eigenvectors, so they define eight different rotation matrices $\{\pm\tilde{u}_1, \pm\tilde{u}_2, \pm\tilde{u}_3\}$ specifying the same principal axes \cite{GalvezCanton1992}.

Although it is possible to reduce this ambiguity to four combinations by only keeping right-handed coordinate systems \cite{GalvezCanton1992}, a heuristic approach is still needed to obtain a unique standard rotation matrix. In our work, we will use the invariants to locally compare 3D image surfaces. For this reason, we first locate points on the surface of the 3D image. Then, for each local patch around these points, a direction vector can be specified using the vertex normal at the surface point, pointing away from the surface. Among the eight rotations, the unique rotation matrix $\tilde{R}$ can be chosen, for example, as the one rotating the vertex normals to the octant in which $x$, $y$, and $z$ coordinates are all nonpositive. Once $\tilde{R}$ is determined, we can define geometric moments of $\tilde{f}(x,y,z)$, which are invariant under rotation, translation, and scaling by
\begin{equation}
	\begin{split}
	    \tilde{\nu}_{ijk} = & \, (\tilde{\mathrm{M}}_{000})^{-\frac{i+j+k}{3}-1} \sum_{x=0}^{N} \sum_{y=0}^{M} \sum_{z=0}^{L} \,\tilde{f}(x,y,z) \\
			\; \cdot   \,(\tilde{\phi}_1&(x,y,z)) ^i  \cdot  ( \tilde{\phi}_2(x,y,z) )^j \cdot  ( \tilde{\phi}_3(x,y,z) )^k, 
	\end{split}
	\label{eqn:tilde_nu_nml}
\end{equation}
where $ \tilde{\phi}_1(x,y,z) = \tilde{R}_{11}(x-\tilde{x}) + \tilde{R}_{12}(y-\tilde{y}) + \tilde{R}_{13}(z-\tilde{z})$, $\tilde{\phi}_2(x,y,z) = \tilde{R}_{21}(x-\tilde{x}) + \tilde{R}_{22}(y-\tilde{y}) + \tilde{R}_{23}(z-\tilde{z})$, and $\tilde{\phi}_3(x,y,z)=\tilde{R}_{31}(x-\tilde{x}) + \tilde{R}_{32}(y-\tilde{y}) + \tilde{R}_{33}(z-\tilde{z})$.

Fig. \ref{fig:weights} shows 2D views of the square root of the 3D weight function $W(x,y,z;\mathbf{p},\mathcal{N})$ in (\ref{eqn:3DWeightFunction}) for $\mathcal{N}=(200,200,200)$ and five different $\mathbf{p}=(p_x,p_y,p_z)$ triplets. To obtain the 2D views, we flatten the function by summing slices along each of the three dimensions.  Note that the coverage of each function differs as the parameters $\mathbf{p}=(p_x,p_y,p_z)$ change. The coverage is the largest for $\mathbf{p}=(0.5,0.5,0.5)$ and becomes smaller when $\mathbf{p}$ approaches faces of the grid. Different weight functions result in loss of translation and scale invariance. One way to overcome this problem is to determine a unique suitable $\mathbf{p}$ triplet, say $\mathbf{p}^{\ast}=(p_x^{\ast},p_y^{\ast},p_z^{\ast})$, and use the corresponding weight $W(x,y,z;\mathbf{p}^{\ast},\mathcal{N})$ for every local region-of-interest by shifting the graph of $W$ to that location, in other words, use the translated weight $W^{\ast}(x,y,z;\mathbf{p}^{\ast},\mathcal{N})=W(x^{\ast},y^{\ast},z^{\ast};\mathbf{p},\mathcal{N})$ with $x^{\ast}=x-Np_x^{\ast}+Np_x$, $y^{\ast}=y-Mp_y^{\ast}+Mp_y$, and $z^{\ast}=z-Lp_z^{\ast}+Lp_z$. Whenever $(x^{\ast}, y^{\ast},z^{\ast})$ is situated outside the grid, we set $W^{\ast}(x,y,z;\mathbf{p}^{\ast},\mathcal{N})=0$. From now on in this paper, we will set $\mathbf{p}^{\ast}=(0.5,0.5,0.5)$ due to the largest coverage and round shape of the corresponding weight function, which is also critical for rotational invariance. In order to preserve the round shape of the weight function, we will always use a cubical grid, i.e., $N=M=L$ or set $\mathcal{N}=(N,N,N)$. Hence, the center of the grid will be at $C=(N/2,N/2,N/2)$. In order to shift the centroid of the auxiliary image $\tilde{f}(x,y,z)$ to the grid center $C$, $\tilde{\nu}_{ijk}$ in (\ref{eqn:tilde_nu_nml}) is modified to

\begin{equation}
  \begin{split}
	 \tilde{\lambda}_{ijk} = & \, (\tilde{\mathrm{M}}_{000})^{-1} \sum_{x=0}^{N} \sum_{y=0}^{N} \sum_{z=0}^{N} \, \tilde{f}(x,y,z) \\
	 \cdot & \,( \,  \tilde{\phi}_1(x,y,z)  / (\tilde{\mathrm{M}}_{000})^{1/3} \,+\, N/2 \, ) ^i \\
	 \cdot & \,( \,  \tilde{\phi}_2(x,y,z)  / (\tilde{\mathrm{M}}_{000})^{1/3} \,+\, N/2 \, ) ^j \\
	 \cdot & \,( \,  \tilde{\phi}_3(x,y,z)   / (\tilde{\mathrm{M}}_{000})^{1/3} \,+\, N/2 \, ) ^k.
  \end{split}
\label{eqn:tilde_nu_nml_modif}
\end{equation}
Using the binomial expansion, $\tilde{\lambda}_{ijk}$ can be written as

\begin{equation}
  \begin{split}
    \tilde{\lambda}_{ijk}  & = \, (\tilde{\mathrm{M}}_{000})^{-1} \sum_{x=0}^{N} \sum_{y=0}^{N} \sum_{z=0}^{N} \tilde{f}(x,y,z) \\
     \cdot & \sum_{r=0}^{i} {\binom{i}{r}} \left(\tilde{\phi}_1(x,y,z)/(\tilde{\mathrm{M}}_{000})^{1/3}\right)^r (N/2)^{i-r}  \\
     \cdot & \sum_{s=0}^{j} {\binom{j}{s}} \left(\tilde{\phi}_2(x,y,z)/(\tilde{\mathrm{M}}_{000})^{1/3}\right)^s (N/2)^{j-s}  \\
     \cdot & \sum_{t=0}^{k} {\binom{k}{t}} \left(\tilde{\phi}_3(x,y,z)/(\tilde{\mathrm{M}}_{000})^{1/3}\right)^t (N/2)^{k-t}.
  \end{split}
  \label{eqn:lin_comb_of_invs_part1}
\end{equation}
which can then be rearranged to
\begin{equation}
  \begin{split}
    \tilde{\lambda}_{ijk}  & = \, \sum_{r=0}^{i} \sum_{s=0}^{j} \sum_{t=0}^{k} {\binom{i}{r}} {\binom{j}{s}} {\binom{k}{t}} \, (N/2)^{i+j+k-r-s-t} \\
     &\cdot\,  (\tilde{\mathrm{M}}_{000})^{-\frac{r+s+t}{3}-1} \sum_{x=0}^{N} \sum_{y=0}^{N} \sum_{z=0}^{N} \tilde{f}(x,y,z) \\
     & \cdot \, (\tilde{\phi}_1(x,y,z))^r \cdot (\tilde{\phi}_2(x,y,z))^s \cdot (\tilde{\phi}_3(x,y,z))^t.
  \end{split}
  \label{eqn:lin_comb_of_invs_part2}
\end{equation}
Thus, 
\begin{equation}
  \begin{split}
    \tilde{\lambda}_{ijk}  = & \, \sum_{r=0}^{i} \sum_{s=0}^{j} \sum_{t=0}^{k} {\binom{i}{r}} {\binom{j}{s}} {\binom{k}{t}} \, (N/2)^{i+j+k-r-s-t} \, \tilde{\nu}_{rst}
  \end{split}
  \label{eqn:lin_comb_of_invs_part3}
\end{equation}
is a linear combination of invariants $\tilde{\nu}_{rst}$ where
\begin{equation}
	\begin{split}
	    \tilde{\nu}_{rst} = & \, (\tilde{\mathrm{M}}_{000})^{-\frac{r+s+t}{3}-1} \sum_{x=0}^{N} \sum_{y=0}^{N} \sum_{z=0}^{N} \,\tilde{f}(x,y,z) \\
			\; \cdot   \,(\tilde{\phi}_1&(x,y,z)) ^r  \cdot  ( \tilde{\phi}_2(x,y,z) )^r \cdot  ( \tilde{\phi}_3(x,y,z) )^t, 
	\end{split}
	\label{eqn:tilde_nu_rst}
\end{equation}
for $r=0,\ldots,i$, $s=0,\ldots,j$ and $t=0,\ldots,k$. Therefore, these new geometric moments are rotation, translation, and scale invariant, and yet centered at the point $(N/2,N/2,N/2)$. If we set $p_x=0.5$, $p_y=0.5$, and $p_z=0.5$ in (\ref{eqn:Qbar}) and replace $\tilde{\mathrm{M}}_{ijk}$ by their invariant counterparts $\tilde{\lambda}_{ijk}$ from (\ref{eqn:tilde_nu_nml_modif}), we obtain a new set of moments which are invariant under rotation, translation, and scaling, i.e.,
\begin{equation}
	\begin{split}
		\tilde{Q}_{nml} = & \, [\Omega (n,m,l;\mathbf{p}^{\ast},\mathcal{N})]^{-1/2} \\
		                  \cdot \sum_{i=0}^{n} &\sum_{j=0}^{m} \sum_{k=0}^{l} \, a_{i,n,0.5,N} \, a_{j,m,0.5,N} \, a_{k,l,0.5,N} \,   \tilde{\lambda}_{ijk}.
	\end{split}
	\label{eqn:local_Kraw-based_inv}
\end{equation}
where $\mathbf{p}^{\ast}=(0.5, 0.5, 0.5)$ and $\mathcal{N}=(N,N,N)$.
This new set of moments will be called {\em 3D Krawtchouk descriptors} and referred as 3DKD in the rest of the paper. Note that 3DKD still depend on the number $N$, so it is important to use the same grid while performing the local comparison of 3D images.

\begin{figure*}[!ht]
	\centering
		\begin{tabular}{cccc}
			\includegraphics[scale=0.36,trim = 2.2cm 2.2cm 2.2cm 2.2cm,clip=true]{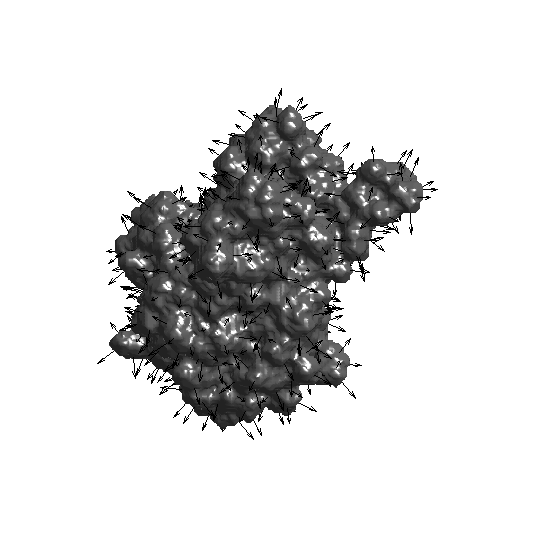}  &
			\includegraphics[scale=0.36,trim = 2.2cm 2.2cm 2.2cm 2.2cm,clip=true]{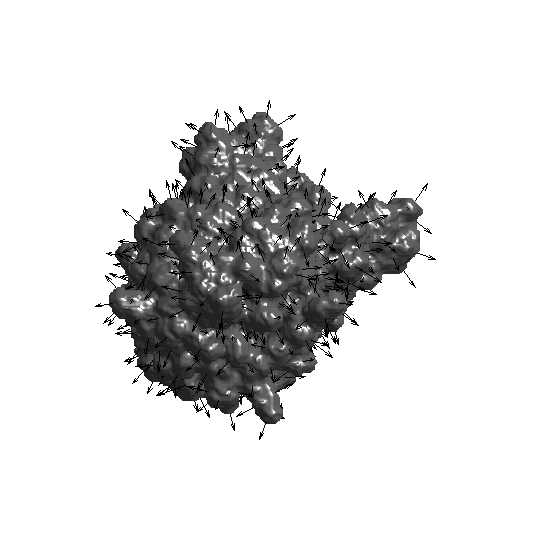}  &
			\includegraphics[scale=0.36,trim = 2.2cm 2.2cm 2.2cm 2.2cm,clip=true]{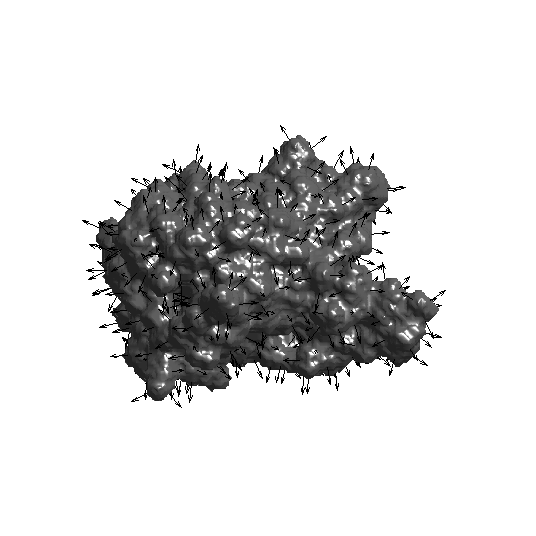}  &
		 	\includegraphics[scale=0.36,trim = 2.2cm 2.2cm 2.2cm 2.2cm,clip=true]{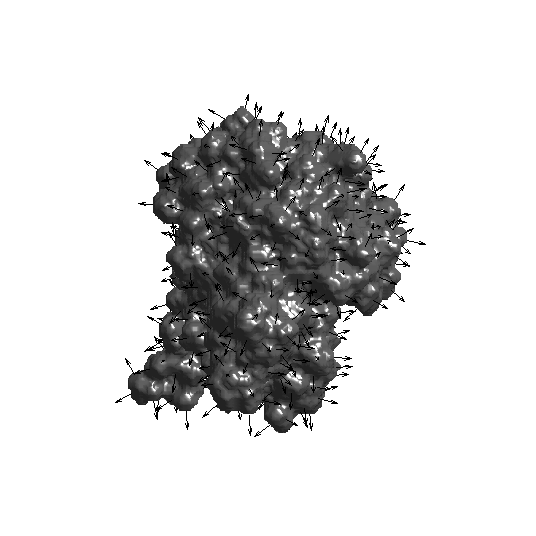} 
		\end{tabular}
		\caption{Query protein surface (1gco.pdb, left) and three target surfaces obtained from the query protein by rotating it using the rotation matrices $S_1$, $S_2$, and $S_3$ in (\ref{eqn:rotations}), respectively. The vertex normals on each surface are also demonstrated.}
		\label{fig:1gco}
\end{figure*}

\section{Computation of Descriptors}
\label{sec:Computation}

The descriptors defined in (\ref{eqn:local_Kraw-based_inv}) requires precomputation of $\tilde{\lambda}_{ijk}$ in (\ref{eqn:lin_comb_of_invs_part3}) which is a linear combination of geometric moments $\tilde{\nu}_{rst}$ given by (\ref{eqn:tilde_nu_rst}). Notice that computation of $\tilde{\nu}_{rst}$ requires exponentiation of three 3D functions $\tilde{\phi}_1$, $\tilde{\phi}_2$, and $\tilde{\phi}_3$, and then element-wise multiplication of four 3D functions, and finally summation over three variables. Thus, direct computation in (\ref{eqn:tilde_nu_rst}) of $\tilde{\nu}_{rst}$ can be quite time-consuming, especially when the grid size $\mathcal{N}=(N,N,N)$ is large. To save computational time, we will first separate the $z$ variable from the $x$ and $y$ variables by rewriting $(\tilde{\phi}_1(x,y,z))^r$, $(\tilde{\phi}_2(x,y,z))^s$, and $(\tilde{\phi}_3(x,y,z))^t$ using the binomial expansion as follows:
\begin{equation}
  	\begin{split}
    (\tilde{\phi}_1(x,y,z))^r = & \,\sum_{\varepsilon_1=0}^{r} \, {\binom{r}{\varepsilon_1}} (\tilde{A}_1(x,y))^{r-\varepsilon_1} \,(\tilde{D}_1(z))^{\varepsilon_1}, \,
	\end{split}
\end{equation}
\begin{equation}
  	\begin{split}
    (\tilde{\phi}_2(x,y,z))^s = & \,\sum_{\varepsilon_2=0}^{s} \, {\binom{s}{\varepsilon_2}} (\tilde{A}_2(x,y))^{s-\varepsilon_2} \,(\tilde{D}_2(z))^{\varepsilon_2}, \,
	\end{split}
\end{equation}
\begin{equation}
  	\begin{split}
    (\tilde{\phi}_3(x,y,z))^t = & \,\sum_{\varepsilon_3=0}^{t} \, {\binom{t}{\varepsilon_3}} (\tilde{A}_3(x,y))^{t-\varepsilon_3} \,(\tilde{D}_3(z))^{\varepsilon_3}, \,
  \end{split}
\end{equation}
where $ \tilde{A}_{\tau}(x,y) = \tilde{R}_{{\tau} 1}(x-\tilde{x}) + \tilde{R}_{{\tau} 2}(y-\tilde{y})$ and $\tilde{D}_{\tau}(z) = \tilde{R}_{{\tau} 3}(z-\tilde{z})$ with ${\tau}=1,2,3$.

Hence, given a voxelized 3D surface function $f(x,y,z)$, a triplet $\mathbf{p}=(p_x,p_y,p_z)$ corresponding to a point on the surface, and the surface normal at that point, an efficient computation of the geometric moments $\tilde{\nu}_{rst}$ in (\ref{eqn:tilde_nu_rst}) can be performed in the following steps.

\begin{enumerate}
	\item Compute the auxiliary image $\tilde{f}(x,y,z)$ using (\ref{eqn:WeightedFunction}).
    
    \item Compute the function
    	\begin{equation}
  			\begin{split}
    			& \tilde{T}(x,y;\varepsilon_1,\varepsilon_2,\varepsilon_3) \\
                = & \,\sum_{z=0}^{N} \, \tilde{f}(x,y,z) \, (\tilde{D}_1(z))^{\varepsilon_1}\,(\tilde{D}_2(z))^{\varepsilon_2}\,(\tilde{D}_3(z))^{\varepsilon_3}
  			\end{split}
		\end{equation}
       	for $0\leq \varepsilon_1+\varepsilon_2+\varepsilon_3 \leq 5$. Note that the $z$ variable is eliminated in this step and the rest of the computations will be carried out in the $x$ and $y$ variables only.
        
	\item Compute 
    	\begin{equation}
  			\begin{split}
    			& \tilde{T}_3(x,y;\varepsilon_1,\varepsilon_2,t) \\
                = & \,\sum_{\varepsilon_3=0}^{t} \, {\binom{t}{\varepsilon_3}}\, (\tilde{A}_3(x,y))^{t-\varepsilon_3}\,\tilde{T}(x,y;\varepsilon_1,\varepsilon_2,\varepsilon_3)
  			\end{split}
		\end{equation}
       	for $0\leq \varepsilon_1+\varepsilon_2+t \leq 5$.
        
	\item Compute 
    	\begin{equation}
  			\begin{split}
    			& \tilde{T}_2(x,y;\varepsilon_1,s,t) \\
                = & \,\sum_{\varepsilon_2=0}^{s} \, {\binom{s}{\varepsilon_2}}\, (\tilde{A}_2(x,y))^{s-\varepsilon_2}\,\tilde{T}_3(x,y;\varepsilon_1,\varepsilon_2,t)
  			\end{split}
		\end{equation}
       	for $0\leq \varepsilon_1+s+t \leq 5$.
        
	\item Compute 
    	\begin{equation}
  			\begin{split}
    			& \tilde{T}_1(x,y;r,s,t) \\
                = & \,\sum_{\varepsilon_1=0}^{r} \, {\binom{r}{\varepsilon_1}}\, (\tilde{A}_1(x,y))^{r-\varepsilon_1}\,\tilde{T}_2(x,y;\varepsilon_1,s,t)
  			\end{split}
		\end{equation}
       	for $0\leq r+s+t \leq 5$.

	\item Compute 
    	\begin{equation}
  			\begin{split}
    			\tilde{\nu}_{rst} = & \,(\tilde{\mathrm{M}}_{000})^{-\frac{r+s+t}{3}-1}\,\sum_{x=0}^{N} \sum_{y=0}^{N} \, \tilde{T}_1(x,y;r,s,t)
  			\end{split}
		\end{equation}
       	for $0\leq r+s+t \leq 5$.
        
\end{enumerate}

Finally, we perform another couple of steps to compute 3DKD of the order up to $5$.

\begin{enumerate}
	\setcounter{enumi}{6}
	\item Compute $\tilde{\lambda}_{ijk}$ in (\ref{eqn:lin_comb_of_invs_part3}) for $0\leq i+j+k\leq 5$ using $\tilde{\nu}_{rst}$ from step 6.
    \item Compute $\tilde{Q}_{nml}$ in (\ref{eqn:local_Kraw-based_inv}) for $0\leq n+m+l\leq 5$ using $\tilde{\lambda}_{ijk}$ from step 7.
\end{enumerate}

Separating the $z$ variable as described above makes the computations very efficient. The computational performance of the algorithm will be shown at the end of Section~\ref{sec:Results}.

\begin{table}[!ht]
\centering
\caption{3DKD Used in Local Comparison}
\footnotesize
\begin{tabular}{|c|c|c|}
	\hline
	 Feature vector ($\varv$) & Size of the vector   & Actual size ($T$)  \\
	\hline
	$K_3$ & 20 & 13  \\
	\hline
	$K_4$ & 35 & 28  \\
	\hline
	$K_5$ & 56 & 49  \\
	\hline
\end{tabular}
\label{table:3DZD_size}
\end{table}

\section{Results and Discussion}
\label{sec:Results}

In this section, we test the local discriminative performance of 3DKD. We use three feature vectors of descriptors
 \begin{equation}
   \begin{split}
     K_3  \, & = \, \{\tilde{Q}_{nml}: 0\leq n+m+l\leq 3\} \\
     K_4  \, & = \, \{\tilde{Q}_{nml}: 0\leq n+m+l\leq 4\} \\
     K_5  \, & = \, \{\tilde{Q}_{nml}: 0\leq n+m+l\leq 5\} \\
   \end{split}
   \label{eqn:3DKDs_used}
 \end{equation}
namely, the descriptors of order up to $3$, $4$, and $5$, that are computed using the algorithm summarized in Section \ref{sec:Computation}. The number of elements in $K_3$, $K_4$, and $K_5$ is 20, 35, and 56, respectively. The seven descriptors ($\tilde{Q}_{000}$, $\tilde{Q}_{100}$, $\tilde{Q}_{010}$, $\tilde{Q}_{001}$, $\tilde{Q}_{011}$, $\tilde{Q}_{101}$, $\tilde{Q}_{110}$) involved in the normalization process were removed, because they take a constant value irrespective of the 3D patch we are working with. The numbers of descriptors used in local comparison are given in Table~\ref{table:3DZD_size}.

As the similarity measure, we use the (squared) Euclidean distance between feature vectors of the same size, namely
\begin{equation}
  \begin{split}
    d(\varv^q,\varv^t) = & \, \sum_{i=1}^{T} (\varv_i^q - \varv_i^t)^2
  \end{split}
\end{equation}
where $\varv^q$ and $\varv^t$ are the feature vectors for a query and a target object, respectively, to be compared, and $T$ is dimension of the feature vector.

\begin{table}[!ht]
\centering
\caption{Test I - Recognition Accuracies (\%)}
\footnotesize
\begin{tabular}{|c|c|c|c|}
	\hline
	 \multicolumn{4}{|l|}{Target 1}   \\
	\hline
	 3DKD feature vector   & Top 1 & Top 5 & Top 10 \\
	\hline
	$K_3$ & 96.8 & 98.2 & 98.6 \\
	\hline
	$K_4$ & 96.2 & 99.0 & 100.0 \\
	\hline
	$K_5$ & 97.0 & 98.0 & 98.6 \\
	\hline
	\hline
	 \multicolumn{4}{|l|}{Target 2}   \\
	\hline
	 3DKD feature vector   & Top 1 & Top 5 & Top 10 \\
	\hline
	$K_3$ &  95.4 &  97.6 &  98.0 \\
	\hline
	$K_4$ &  93.6 &  98.0 &  98.8 \\
	\hline
	$K_5$ &  95.6 &  97.6 &  98.0 \\
	\hline
		\hline
	 \multicolumn{4}{|l|}{Target 3}   \\
	\hline
	 3DKD feature vector   & Top 1 & Top 5 & Top 10 \\
	\hline
	$K_3$ &  95.8 &  98.4 &  98.6 \\
	\hline
	$K_4$ &  94.8 &  98.4 &  99.0 \\
	\hline
	$K_5$ &  96.0 &  98.2 &  98.6 \\
	\hline
\end{tabular}
\label{table:recog_acc_1}
\end{table}

\subsection{Local Comparison Test I}

We first test 3DKD for comparison of local patches on protein surfaces. We have downloaded the PDB file of a protein (PDB ID: 1GCO) from PDB \cite{PDB2000}, and generated a voxelized surface of the protein using 3D-Surfer \cite{3DSURFER}. Using the GETPOINTS subroutine of LZerD docking suite \cite{LZERD,LZERDSuite}, we have specified 1608 vertex points on the protein surface as well as the normal vectors at these points pointing outside the surface. We have then reduced the number of points to 500 so that the minimal distance between neighboring points is more than 3~\AA~(See Fig.~\ref{fig:1gco}, left). Each of these points and the normal vectors are used to represent the center of a patch on the surface. These normal vectors are also used for determining the unique rotation matrix $R$ as described in Section~\ref{sec:3DKD}. The 3DKD corresponding to each patch is computed and stored as the query dataset. Then, the protein structure and the 500 surface points are rotated so that each patch moves to a different location and orientation (see Fig.~\ref{fig:1gco}). We have used three different rotation matrices $S_1$, $S_2$, and $S_3$ and obtained three target sets. The 3DKD corresponding to each patch in the target sets is computed and stored as Target 1, Target 2, and Target 3. The rotations $S_1$, $S_2$, and $S_3$ are given in (\ref{eqn:rotations}). $S_1$ rotates the query protein $90^{\circ}$ about the $x$-axis, $S_2$ rotates the query protein first $90^{\circ}$ about the $x$-axis, then $45^{\circ}$ about the $y$-axis, and then $30^{\circ}$ about the $z$-axis, and $S_3$ is a randomly generated rotation matrix.

\begin{equation}
  \begin{split}
    S_1 \, = & \, 
    \left[\begin{matrix}
    	1 &  0 & 0 \\ 
    	0 &  0 & 1 \\ 
    	0 & -1 & 0
    \end{matrix}\right] \\
    S_2 \, = & \, 
    \left[\begin{matrix}
    	\sqrt{6}/4 &  \sqrt{6}/4 & 1/2 \\ 
    	-\sqrt{2}/4 &  -\sqrt{2}/4 & \sqrt{3}/2 \\ 
    	\sqrt{2}/2 & -\sqrt{2}/2 & 0
    \end{matrix}\right] \\
    S_3 \, = & \, 
    \left[\begin{matrix}
    	-0.8256 &  0.4039 & -0.3940 \\ 
    	-0.2008 & -0.8629 & -0.4637 \\ 
    	-0.5273 & -0.3037 &  0.7936
    \end{matrix}\right] \\    
  \end{split}
  \label{eqn:rotations}
\end{equation}

Each of the 500 feature vectors in the first set is queried and compared with the 500 feature vectors in a target set. The results are ranked using the Euclidean distance. If the same patch as the query ranks top (Top 1), it is labeled as ``correctly classified''. Otherwise, we look at the top five results in the ranking (Top 5), or the top ten results (Top 10). We have then calculated the recognition accuracies as 
\begin{equation}
    r \,= \dfrac{\,\mbox{Number of ``correctly classified" queries}
\,}{\mbox{Total number of query inputs}}
\end{equation}
whose denominator is equal to $500$.

We have obtained very high recognition accuracies ranging between 93.6\% and 97\% for the Top 1 case (see Table~\ref{table:recog_acc_1}). The recognition accuracies stay between 97.6\% and 99\% for the Top 5 case. For the Top 10 case, it reaches up to between 98\% and 100\%. The results show that the 3DKD's are successful in local patch comparison for this small problem. For the Top 1 case, $K_5$ performs best with all three targets. $K_4$ gives the highest recognition accuracies for all targets when our classification criterion is relaxed to Top 5 or Top 10.

\begin{figure}[!ht]
\centering
		\begin{tabular}{c}
			\includegraphics[scale=0.33,trim = 1.7cm .3cm 1.5cm .6cm,clip=true]{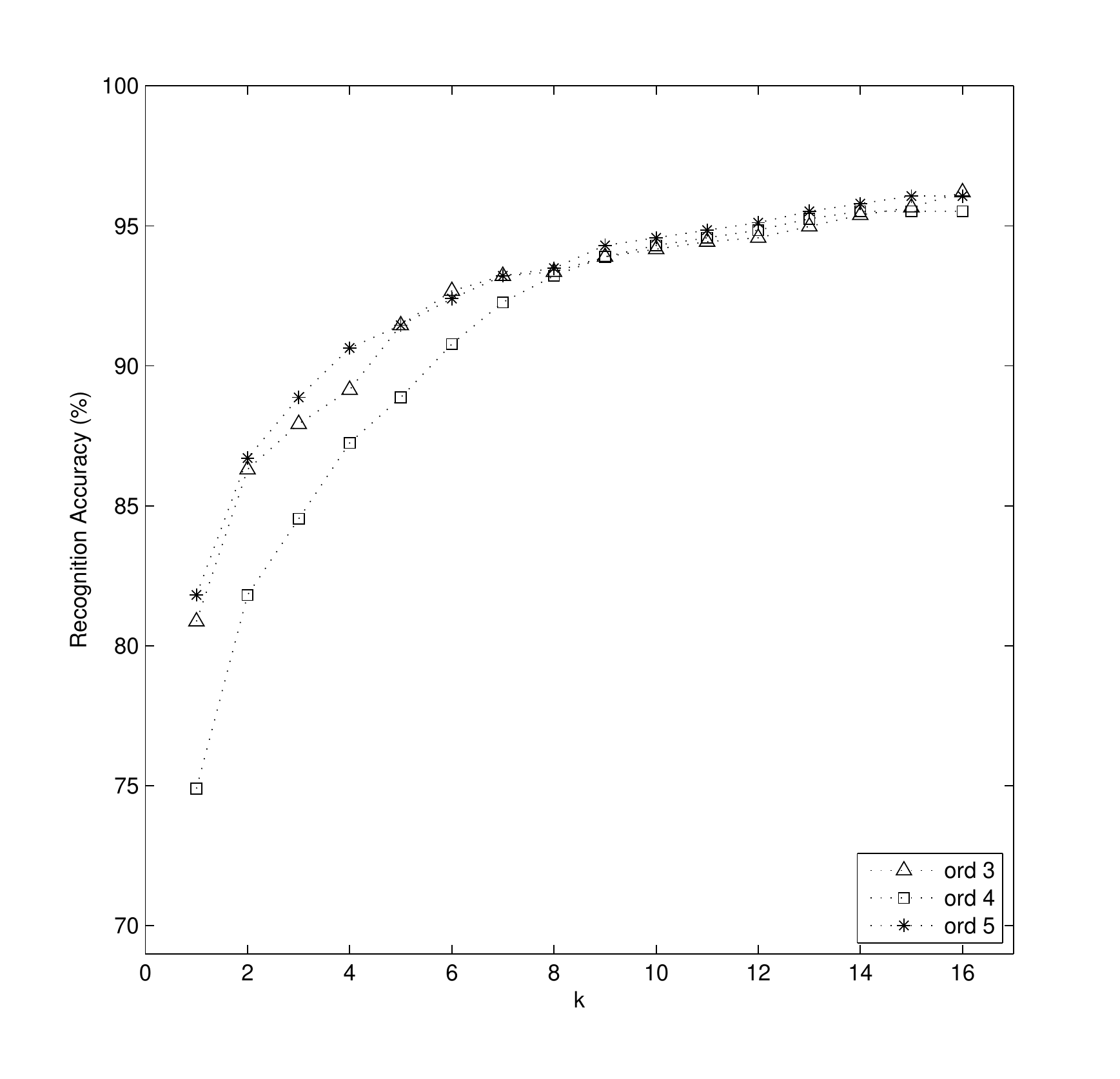} \\
			\includegraphics[scale=0.33,trim = 1.7cm .3cm 1.5cm .6cm,clip=true]{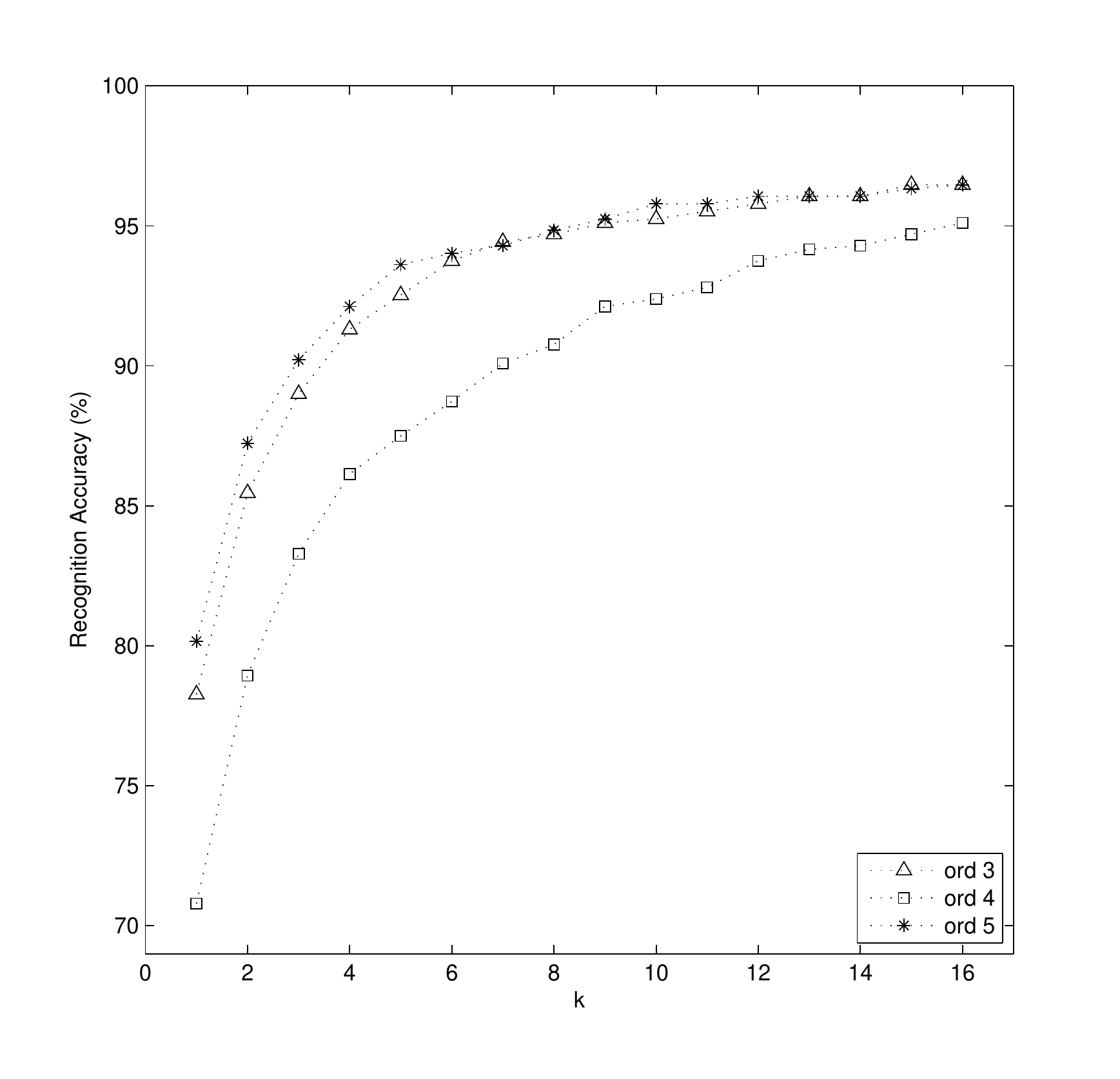} \\
			\includegraphics[scale=0.33,trim = 1.7cm .3cm 1.5cm .6cm,clip=true]{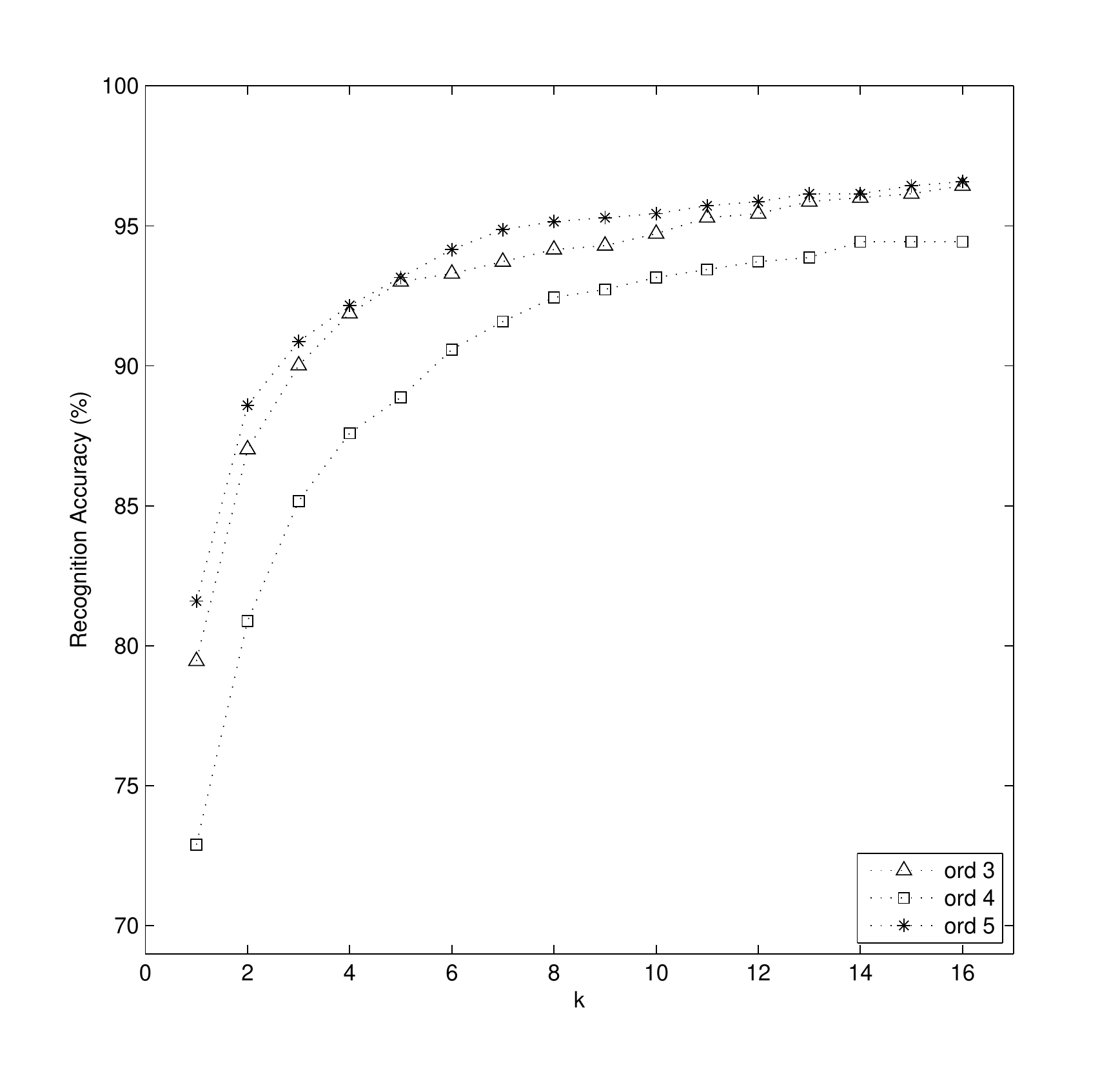}
		\end{tabular}
		\caption{\textbf{Test II - Recognition Accuracies (\%) vs. the number of closest patches in ranking (k) for $S_1$, $S_2$, and $S_3$ from the top to the bottom panel, respectively.}}
		\label{fig:tp2}
\end{figure}

\subsection{Local Comparison Test II}

We also test the local performance of 3DKD on a more difficult problem as follows. The surface grid, vertex points and normal vectors of the query protein, namely 1608 query patches, are generated as before. For the three target sets, the target proteins and their voxelized surface grids are also generated as before. However, the surface points and vertex normals of the target sets are redistributed using GETPOINTS so that we have a new set of points and normal vectors different than the ones obtained before. This occurs due to a randomized subroutine of the program. This time, we have obtained 1557, 1557, and 1546 such points in Target 1, Target 2, and Target 3, respectively. For each point and the corresponding surface patch, the 3DKD's are computed and stored. The number of query patches is then further reduced so that each query patch remains with at least five neighboring target patches when both query and target proteins are considered superimposed. Here, by a neighbor, we mean that the physical distance between centers of the query patch and the target patch is less than 3~\AA. So, the number of query patches to be compared with those in the target sets now depends on each individual target protein, and we obtain 737, 736, and 701 such query patches corresponding to Target 1, Target 2, and Target 3, respectively.

Each of the feature vectors in the query set is compared with those from the corresponding target set. The results are ranked using the Euclidean distance and collected for three target sets. If one of the five neighbors of the query patch ranks within top $k$, it is labeled as ``correctly classified''. We computed recognition accuracies for $k=1,\ldots,16$ where $k=16$ approximately corresponds to the top $1$-percentile.

The results are shown in Fig.~\ref{fig:tp2}. Compared to the results of the Test I (Table~\ref{table:recog_acc_1}), the accuracies dropped about 25 \% points for the $k=1$ results. When we only look at the top result ($k=1$) in the rankings, $K_5$ is the most successful set of descriptors. The performances of $K_3$ and $K_5$ are comparable and higher than $K_4$ for almost all $k$ values in all three cases. In Test II, we have actually tested how each 3DKD vector is tolerant to slight changes in the location of patch centers. From Fig.~\ref{fig:tp2}, it is clear that the descriptors in $K_4$, in particular, the $4$th order descriptors that are not in $K_3$, are quite sensitive to such patch center shifts. This deficiency appears to be corrected in $K_5$ with the addition of $5$th order descriptors.

\subsection{Comparison with 3D Zernike Descriptors}

Next, we compare the discriminative performances of 3DKD and 3D Zernike descriptors (3DZD) \cite{NovotniKlein2003} computed for local protein surface patches. To compute 3DZD's, we used the same query and Target 1 protein from Test I. The 3D Zernike descriptors are not able to extract local features from an object directly as 3DKD's do; yet Zernike functions are defined as continuous functions whose domain is the unit ball. For this reason, we first mapped each patch into the unit ball by considering each surface point as the patch center and placing a sphere of $6$~\AA~radius around each point. $6$~\AA~here corresponds to the size of a typical ligand-binding pocket on a protein surface. The patch cut from the surface is then mapped into the unit ball so that the center of mass of the cropped patch is placed at the coordinate origin. The geometric moments and hence the 3D Zernike moments of the order up to $12$ and $15$ (3DZD\_12 and 3DZD\_15, respectively) are computed for each patch using the algorithm provided in \cite{SitEtAl2013}. Finally, the rotation invariant 3DZD's are computed from these moments using the formula provided in \cite{NovotniKlein2003}. The above work is repeated using a larger sphere for each patch ($9$~\AA~radius) since there are some well-known ligands binding to proteins by forming larger pockets. In order to make the comparison fairer, we have recomputed 3DKDs ($K_5$) using the same surface function but taking on zero value at voxels that remain outside the spheres used above. By this occlusion, we ensure that 3DKDs use the same local information as 3DZDs do.

The recognition accuracies for 3DKDs ($K_5$) and 3DZDs using different order of descriptors and patch radii are shown in Table~\ref{table:comp_with_3DZD}. The results show that our method is clearly better than 3DZDs in local feature extraction, even when 3DZDs are allowed to use more invariants. The Top 1 prediction in our method is 90\%, whereas it is only 7.4\% in 3DZDs. Increasing the order of invariants to $15$ (i.e., the vector size to $72$) does not significantly improve the performance of 3DZD. Both methods give better recognition when the patch size is increased from $6$ to $9$~\AA, but 3DKD outperforms 3DZD in all cases shown in Table~\ref{table:comp_with_3DZD}.

\begin{table}[!ht]
\centering
\caption{Comparison of $K_5$ with 3D Zernike Descriptors (3DZD)}
\footnotesize
\begin{tabular}{|c|c|c|c|c|c|}
	\hline
	 Patch  &  Feature  & Vector  & Top 1 & Top 5 & Top 10 \\
      size &  vector  &  size & \% & \% & \% \\
	\hline
	\hline
	 6~\AA & 3DKD\_$K_5$   & 49 & 90.0 & 97.2 & 98.0 \\
    \cline{2-6}  
	 & 3DZD\_12  & 49 & 7.4  & 20.0 & 29.0 \\
    \cline{2-6} 
	 & 3DZD\_15  & 72 & 8.8 & 18.6 & 28.2 \\
    \hline
    \hline
    9~\AA & 3DKD\_$K_5$   & 49 & 95.6 & 97.8 & 98.6 \\
    \cline{2-6}  
	 & 3DZD\_12  & 49 & 26.0 & 46.2 & 58.0 \\
    \cline{2-6} 
	 & 3DZD\_15  & 72 & 26.4 & 50.4 & 61.8 \\
	\hline
\end{tabular}
\label{table:comp_with_3DZD}
\end{table}

\begin{figure}[!ht]
    \begin{tabular}{c}
      \includegraphics[scale=0.165]{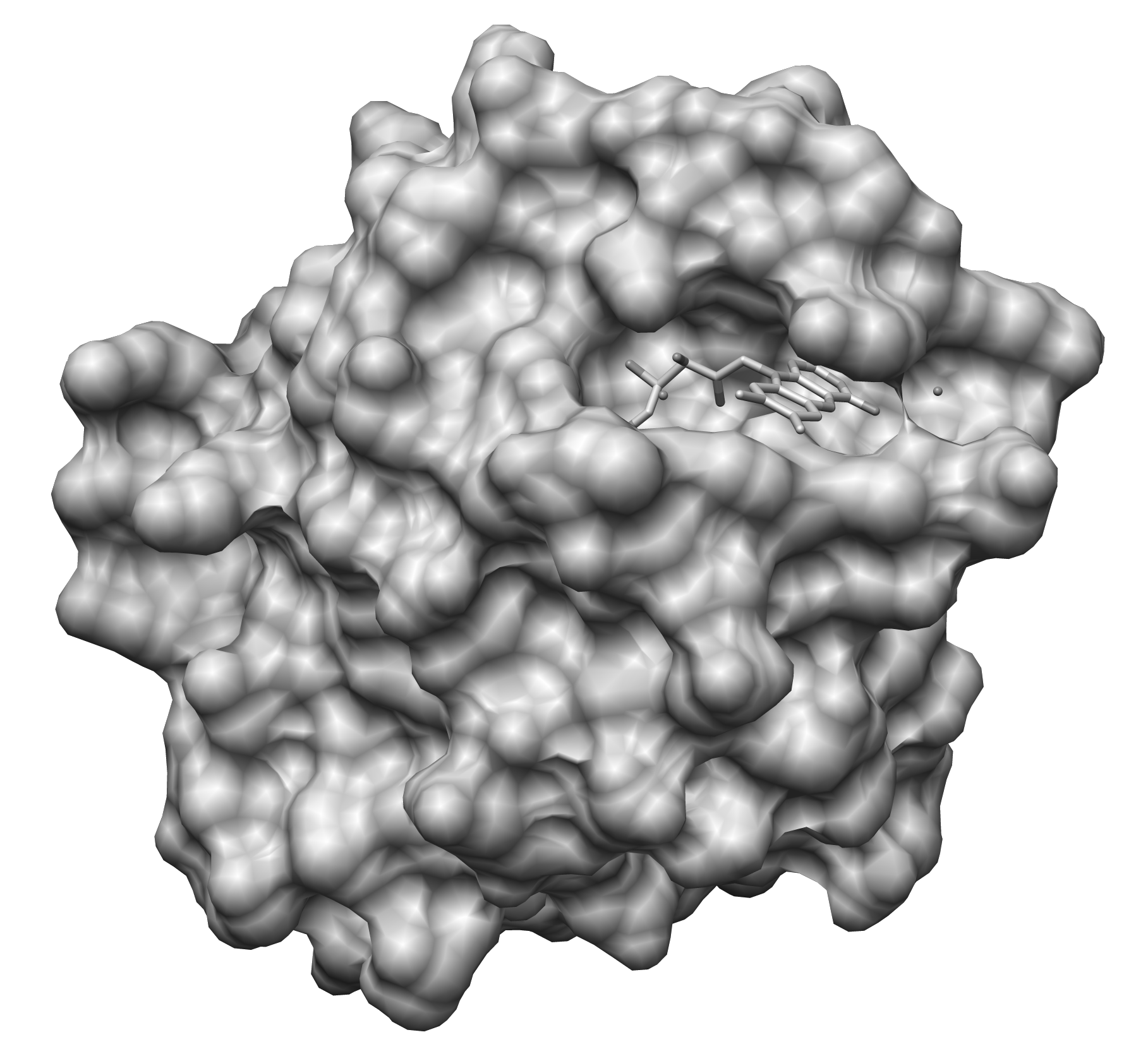} \\ 
      \includegraphics[scale=0.125]{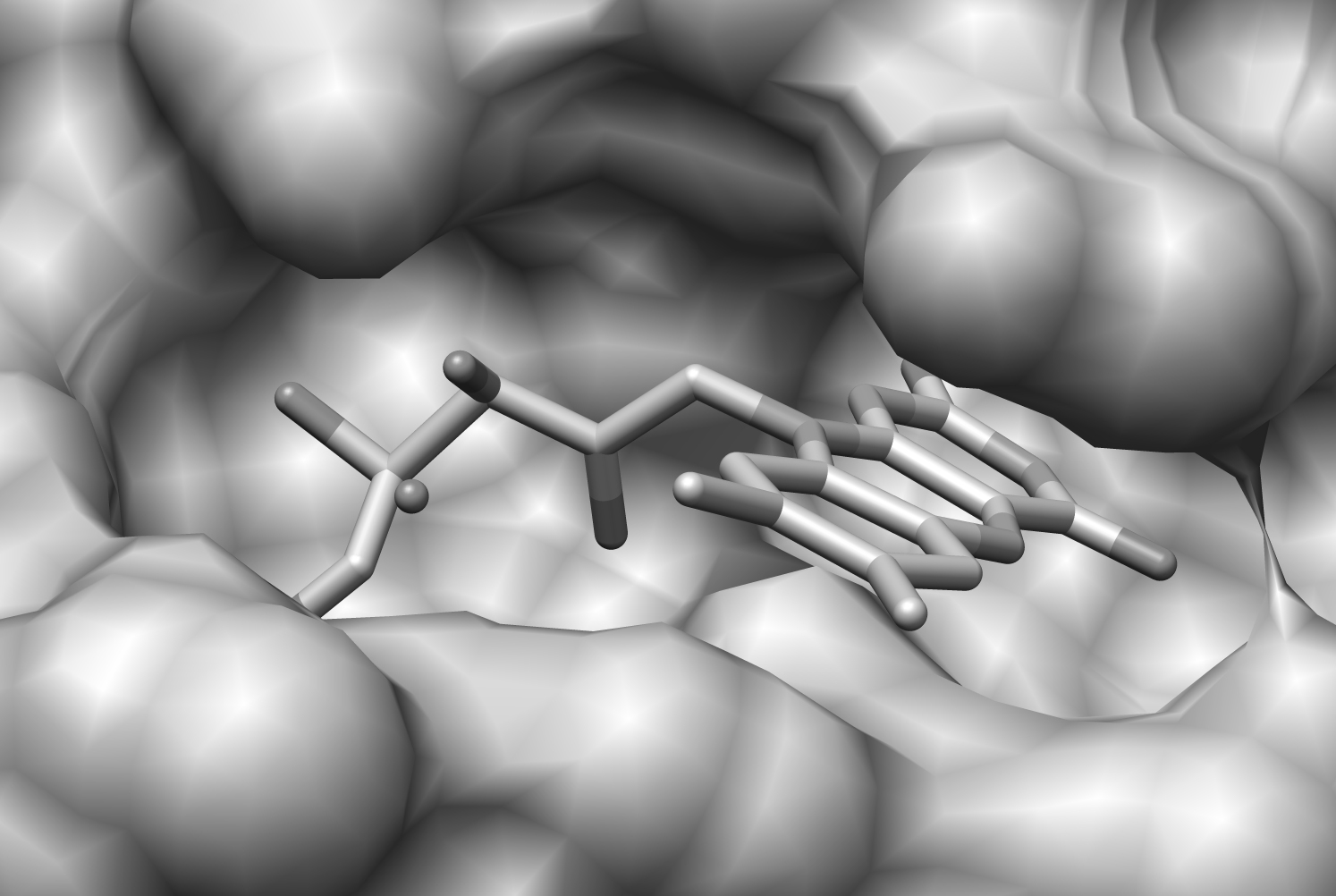} \\ 
    \end{tabular}
    \caption{\footnotesize{An example of a ligand binding pocket on a protein surface. Receptor protein, top: FMN-binding domain of human cytochrome P450 reductase, PDB ID: 1B1C. Binding ligand, bottom: FMN (Flavin mononucleotide). Images were rendered with UCSF Chimera \cite{UCSFChimera}.}}
    \label{fig:binding-ligand-example}
\end{figure}

\begin{figure}[!ht]
    \begin{tabular}{ccc}
    \small{AMP} & \small{ATP} & \small{FAD} \\
      \includegraphics[scale=0.1195]{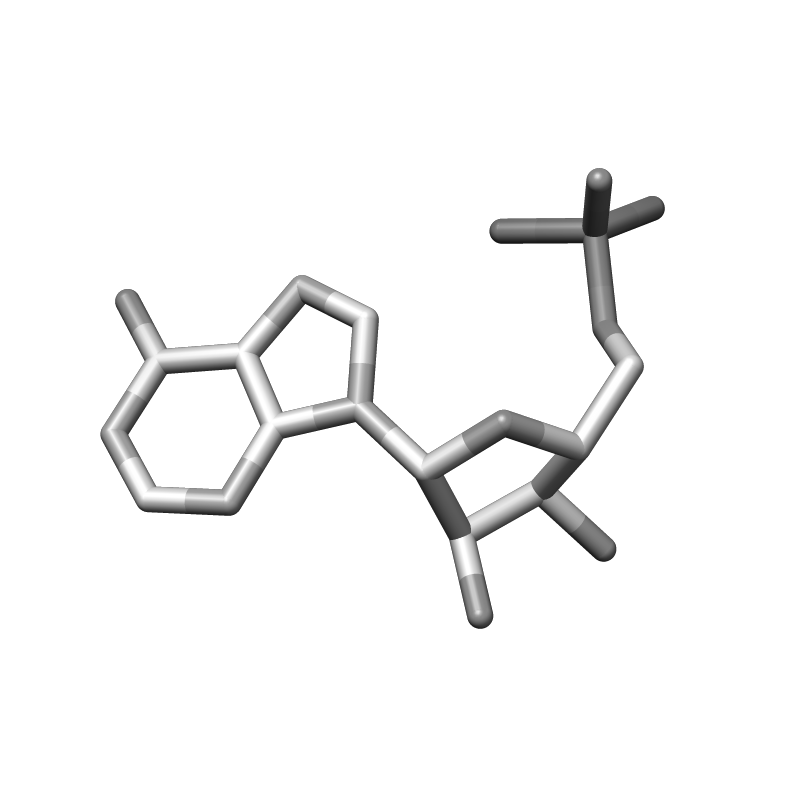} & 
      \includegraphics[scale=0.1195]{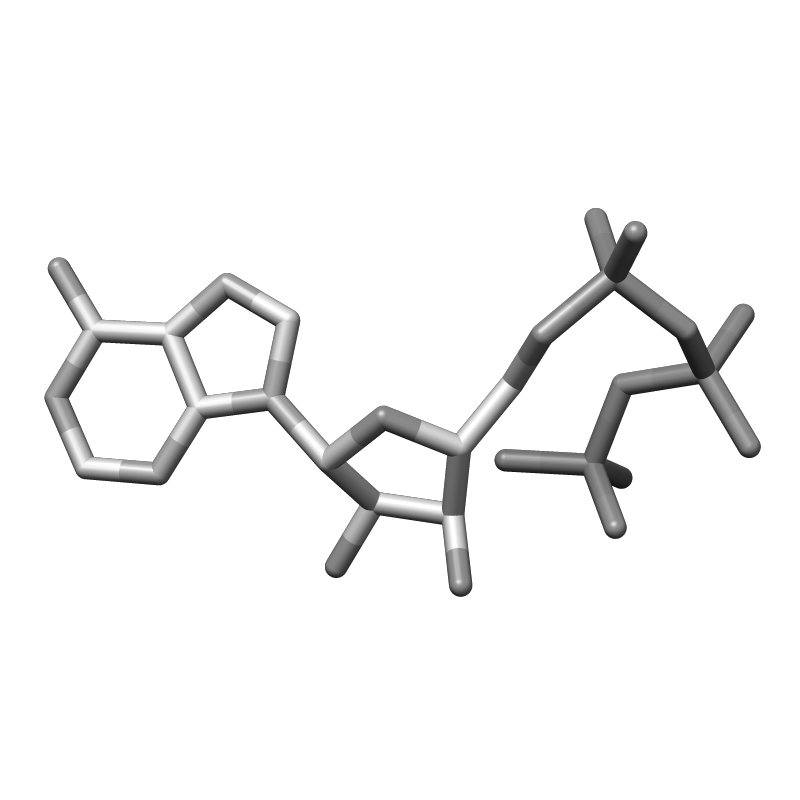} & 
      \includegraphics[scale=0.1195]{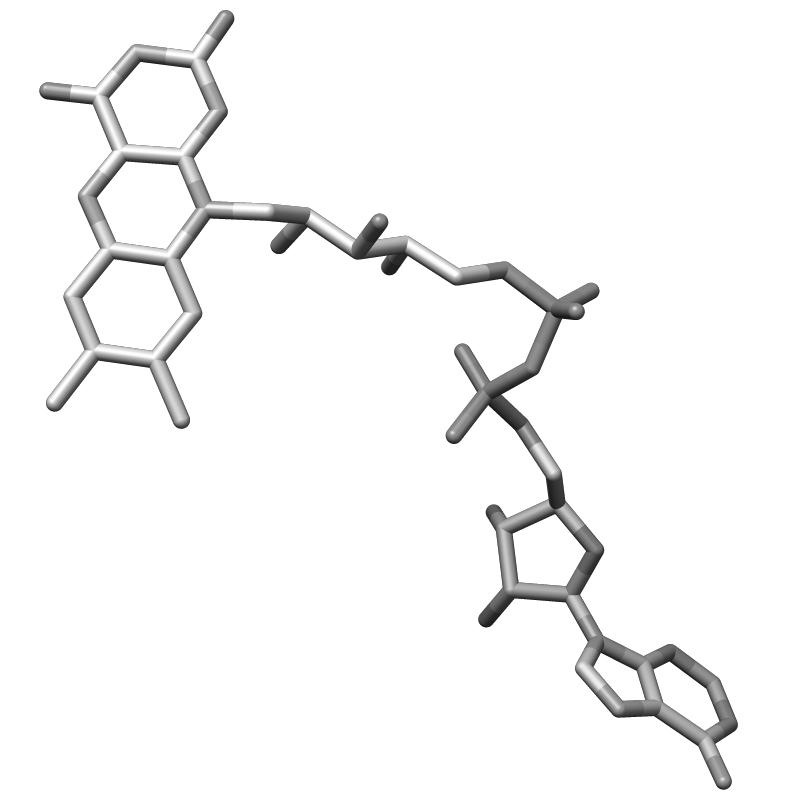} \\ 
	\small{FMN} & \small{FUC} & \small{GAL} \\
      \includegraphics[scale=0.1195]{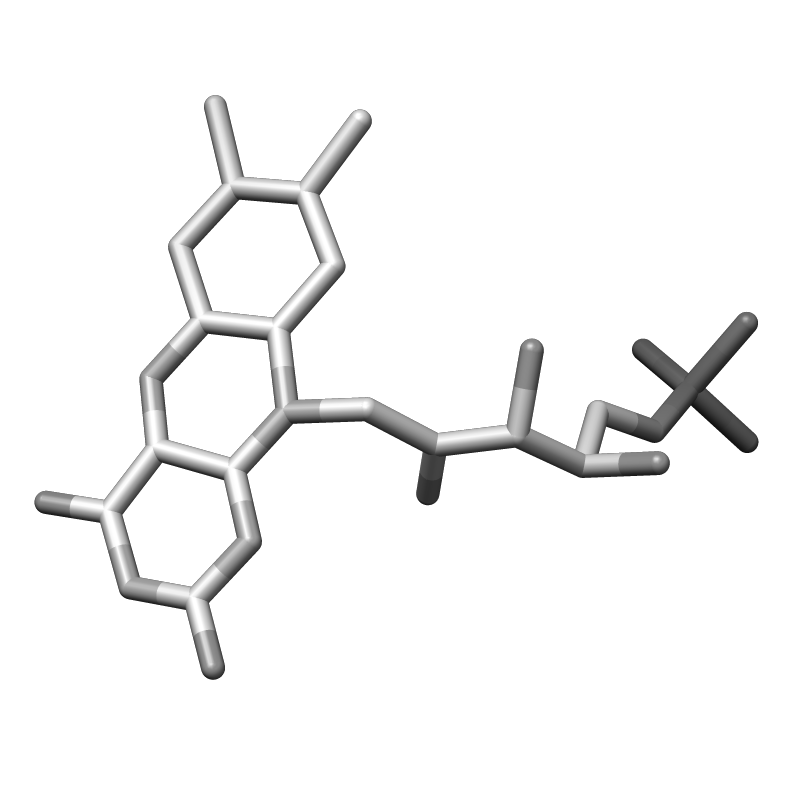} & 
      \includegraphics[scale=0.1195]{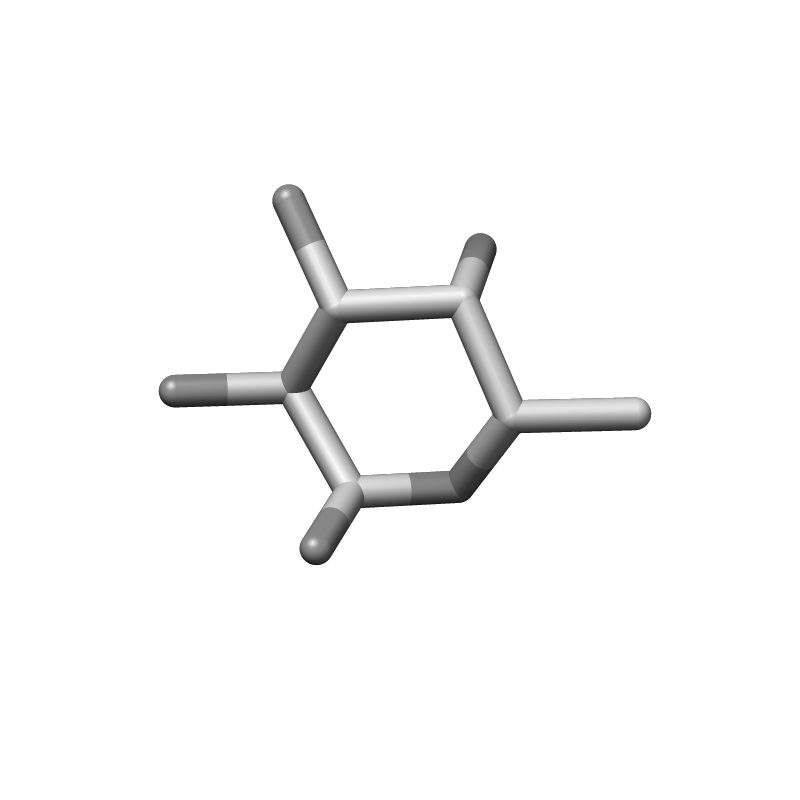} & 
      \includegraphics[scale=0.1195]{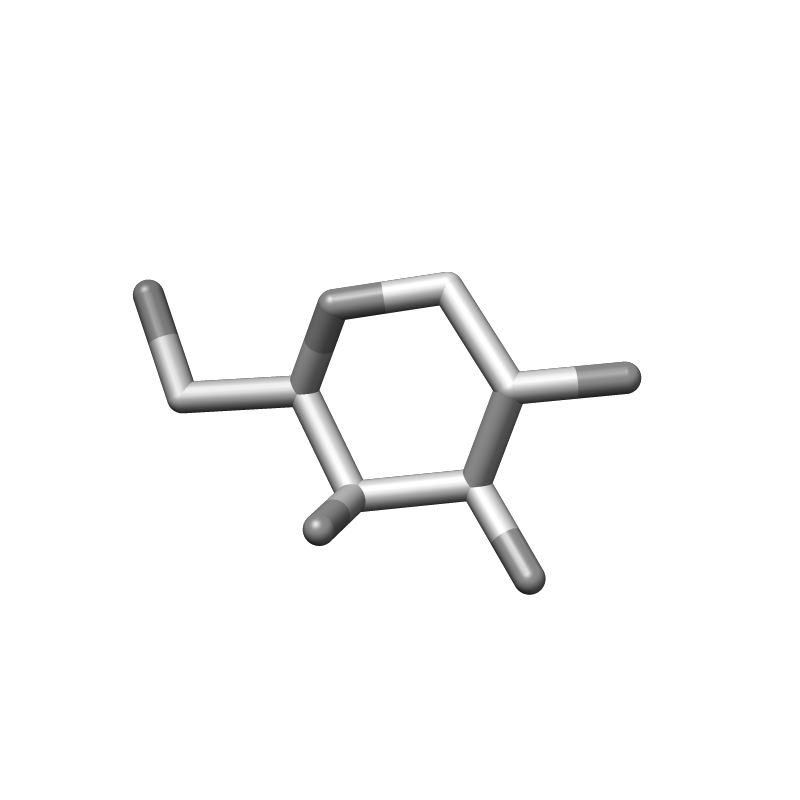} \\
	\small{GLC} & \small{HEM} & \small{MAN} \\
      \includegraphics[scale=0.1195]{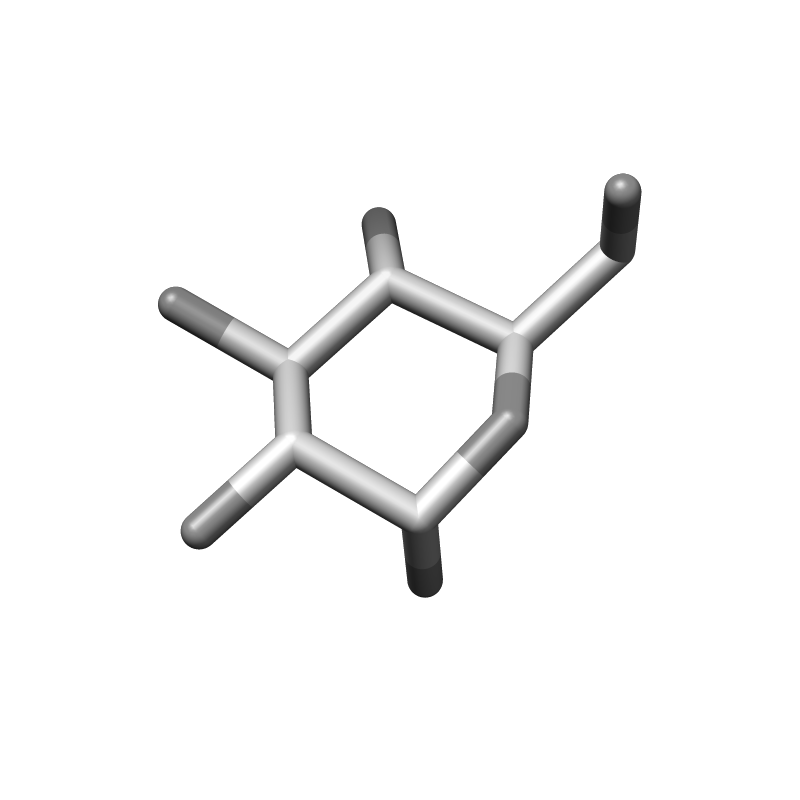} & 
      \includegraphics[scale=0.1195]{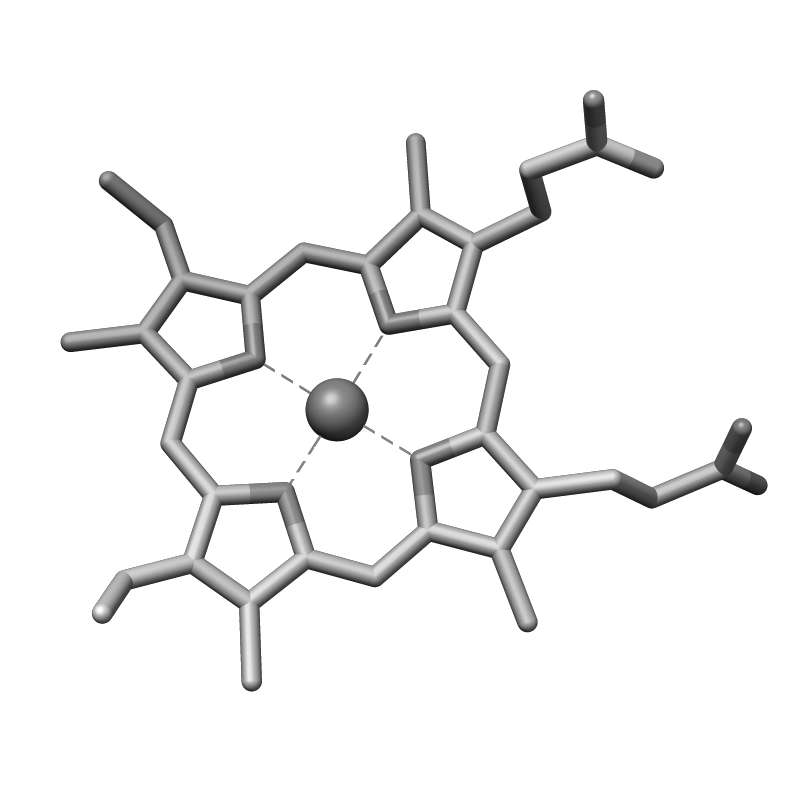} & 
      \includegraphics[scale=0.1195]{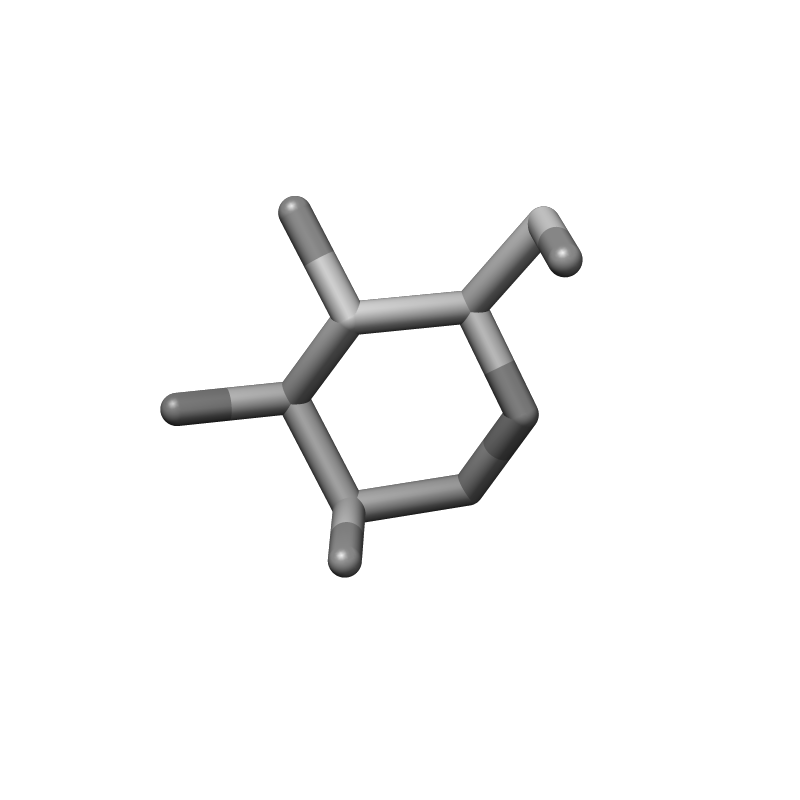} \\
	\small{NAD} & \small{PLM} &  \\
      \includegraphics[scale=0.1195]{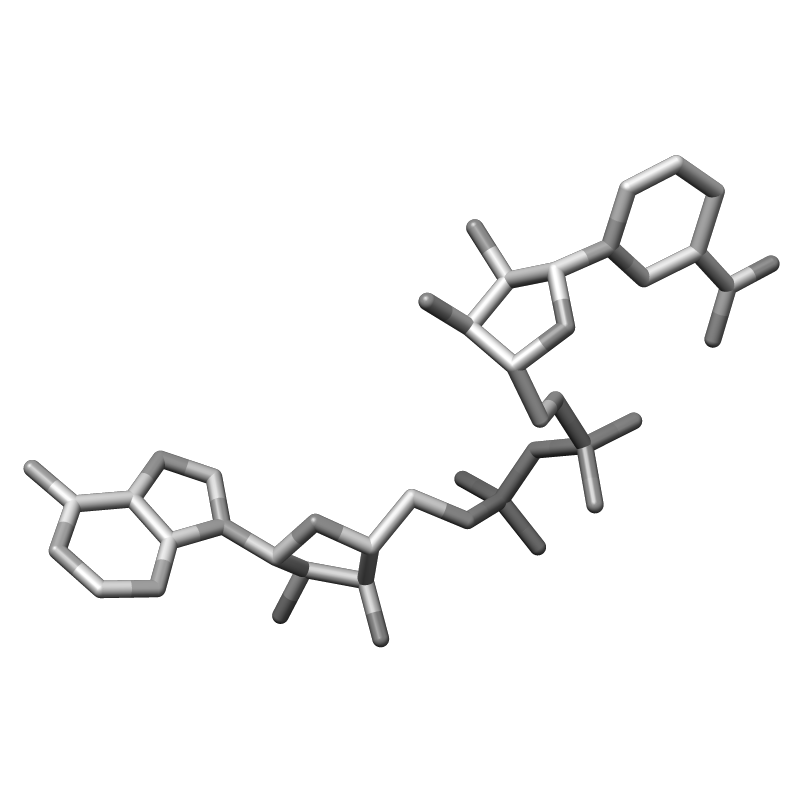} & 
      \includegraphics[scale=0.1195]{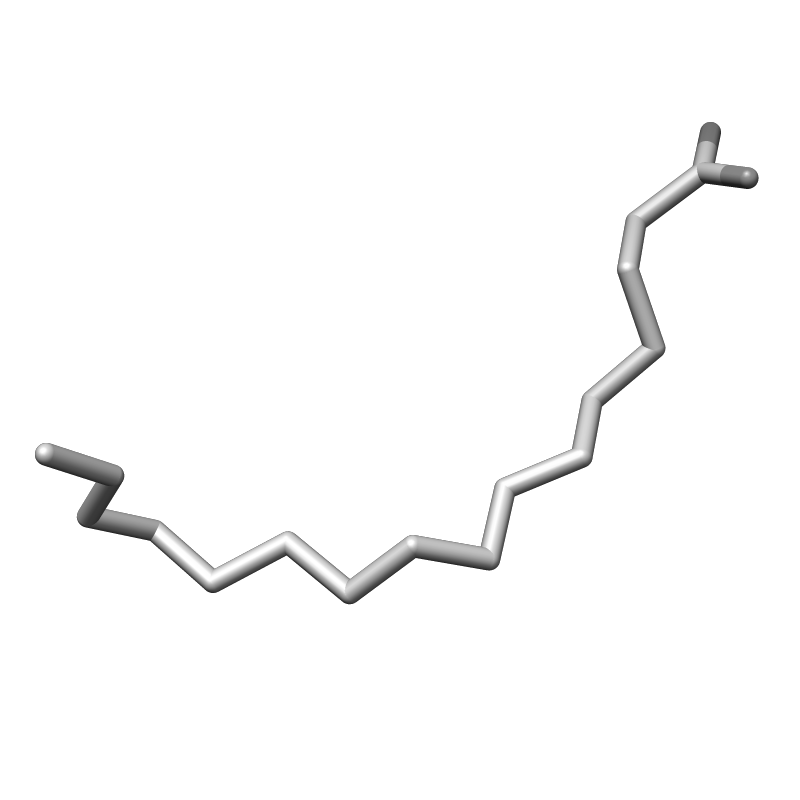} &  \\
    \end{tabular}
    \caption{\footnotesize{Eleven ligand structures known to bound to proteins in the benchmark dataset. The ligand structures are shown for adenosine monophosphate (AMP), adenosine triphosphate (ATP), flavin adenine dinucleotide (FAD), flavin mononucleotide (FMN), fucose (FUC), galactose (GAL), glucose (GLC), heme (HEM), mannose (MAN), nicotinamide adenine dinucleotide (NAD), and palmitic acid (PLM). Images were rendered with UCSF Chimera \cite{UCSFChimera}.}}
    \label{fig:Ligands}
\end{figure}

\subsection{Binding Ligand Prediction}


\begin{table*}
	\centering
	\caption{The Ligand Pocket Benchmark Dataset}
    \label{table:dataset}
	\footnotesize
		\begin{tabular}{|c|rrrrrrrrrrr|}
			\hline
			Binding ligand molecule   & AMP & ATP & FAD & FMN & FUC & GAL & GLC & HEM & MAN & NAD & PLM  \\
			\hline
			Average size (\AA)              & 6.4 & 7.6 & 11.9 & 7.4 & 3.8 & 4.2 & 3.9 & 8.7 & 4.1 & 10.4 & 8.7   \\
			Number of query pockets (497)         &  44 & 44 & 82 & 49 & 7 & 15 & 27 & 146 & 33 &  39 & 11  \\
			Number of patches (11039)        &  619  & 840 & 2663 & 978 & 49 & 125 & 195 & 4089 & 248 & 1095 & 138  \\
			Average number of patches &  14.1  & 19.1 & 32.5 & 20.0 & 7.0 & 8.3 & 7.2 & 28.0 & 7.5 & 28.1 & 12.5   \\
			\hline
		\end{tabular}
\end{table*}


\begin{table*}
	\centering
	\caption{Binding Ligand Prediction Accuracies (\%) Using 3DKD}
    \label{table:prediction_accuracies}
	\footnotesize
		\begin{tabular}{|c|c|c|ccccccccccc|c|}
			\hline
			Rank & Descriptor & $k$ & AMP & ATP & FAD & FMN & FUC & GAL & GLC & HEM & MAN & NAD & PLM & Average  \\
			\hline
			\hline
			Top 1 & $K_3$  &  6 &  11.4 & 20.5 & \textbf{48.8} & 20.4 & \textbf{85.7} & 20.0 & 25.9 & \textbf{91.1} & 45.5 & 33.3 & 18.2 & 38.2  \\
			     & $K_4$  &  2 &  \textbf{15.9} & \textbf{27.3} & 32.9 & \textbf{22.4} & \textbf{85.7} & \textbf{26.7} & \textbf{29.6} & 80.1 & \textbf{48.5} & 30.8 & \textbf{27.3} & 38.8  \\
			     & $K_5$  &  3 &  13.6 & \textbf{27.3} & \textbf{48.8} & \textbf{22.4} & \textbf{85.7} & \textbf{26.7} & 22.2 & 87.0 & \textbf{48.5} & \textbf{41.0} & \textbf{27.3} & \textbf{41.0}  \\
                 \hline
                 & Pocket Surfer & & 0.0 & 21.4 & \textbf{50.0} & 0.0 & 21.4 & \textbf{38.9} & 0.0 & 87.5 & 38.9 & \textbf{60.0} & \textbf{92.3} & 37.3 \\
                 \hline
			     & Random &    &  8.6 & 8.6  & 16.2 &  9.6 &  1.3 &  2.8 &  5.3 & 29.2 &   6.4 &  7.6 &  2.0 &  8.9 \\
			\hline
			\hline
			Top 3 & $K_3$  &  6 &  40.9 & 43.2 & 79.3 & 44.9 & \textbf{85.7} & 46.7 & 48.1 & 96.6 & \textbf{78.8} & 61.5 & 45.5 & 61.0 \\
			     & $K_4$  &  2 &  40.9 & 52.3 & 72.0 & 38.8 & \textbf{85.7} & 40.0 & 63.0 & 95.2 & \textbf{78.8} & 61.5 & 54.5 & 62.1 \\
			     & $K_5$  &  3 &  43.2 & 52.3 & 73.2 & \textbf{46.9} & \textbf{85.7} & 53.3 & 48.1 & 96.6 & \textbf{78.8} & 71.8 & 63.6 & 64.9 \\
                 \hline
                 & Pocket Surfer & & \textbf{77.8} &  \textbf{100.0} &  \textbf{90.0} &  16.7 & \textbf{85.7} & \textbf{80.6} & \textbf{80.0} & \textbf{100.0} & 72.2 & \textbf{100.0} & \textbf{92.3} & \textbf{81.4}  \\
                 \hline
			     & Random &    &  23.8 & 23.9 & 41.4 & 26.3 & 3.7 & 8.4 & 15.2 & 64.8 & 18.1 & 21.5 &  5.8 & 23.0 \\
			\hline
		\end{tabular}
\end{table*}


Finally, we test 3DKD on binding ligand prediction for proteins, which is one of the important tasks in bioinformatics as it addresses a central question in molecular biology, protein function \cite{SealKihara2012,ZhuEtAl2015}, and has real-life application in computational drug design \cite{RosenbergGoldblum2006}. Ligand molecules that bind to a local surface region in a protein can be predicted by finding similar local pockets of known binding ligands in the structure database. An example of a ligand binding pocket is demonstrated in Fig.~\ref{fig:binding-ligand-example}. In order to test 3DKD on binding ligand prediction, we have constructed a benchmark dataset of $463$ proteins already known to bind to $11$ different ligands. See Fig.~\ref{fig:Ligands} for these ligand structures. For each protein in the dataset, surface vertices and normals are generated using the GETPOINTS subroutine as in the previous tests. In the dataset, we also have the PDB file of each protein, which not only contains the coordinates of all atoms in the protein, but also those that belong to the bound ligand. For each atom in the ligand structure, we have selected the nearest surface vertex on the protein and annotated it with the bound ligand type. The collection of all such points and the 3DKDs of the patches around these points are all stored in a `patch database' of $11,039$ patches together with their annotations of binding ligands. Thus, for each query pocket, a database search is performed for each of the patches in the query pocket, and a patch score is assigned to each patch based on the database rankings.

We query total $497$ pockets for this task. See Table \ref{table:dataset} for the number of query pockets for each ligand type. In Table \ref{table:dataset}, we have also listed the number of patches that corresponds to a ligand type. The number of patches associated with each pocket depends on the size of the pocket. By the average size of a pocket, we mean the radius of a sphere that encapsulates all surface vertices annotated with that ligand type. As can be seen from Table \ref{table:dataset}, there is a high correlation between the size of a typical pocket and the average number of patches included in that pocket. For each patch in the query pocket, we compute a patch score based on the following formula:
\begin{equation}
  \begin{split}
     \mbox{Patch\_score}(p,F,k)\, = & \, \sum_{i=1}^{k} \left(\delta_{l(i),F}\log(n/i)\right)\cdot\dfrac{\sum_{i=1}^{k} \delta_{l(i),F}}{\sum_{i=1}^{n} \delta_{l(i),F}},\\   
  \end{split}
  \label{eqn:patch_score}
\end{equation}
where $l(i)$ denotes the ligand type (e.g. AMP, ATP, etc.) of the $i$-th closest patch to the query, $n$ is the number of patches in the patch database, and the function $\delta_{l(i),F}$ is equal to $1$ if $i$-th patch is of type $F$, and $0$ otherwise. This formula is used before as a pocket score for binding-ligand prediction in \cite{SealKihara2012}. The first term in (\ref{eqn:patch_score}) is to only involve $k$ closest patches in the patch database to each patch from the query pocket, assigning a higher score to a patch with a higher rank. The second term is to normalize the score by the number of patches of the same type $F$ included in the patch database so that the results are not biased in favor of highly populated ligand types in the database.

For each of the patches that belong to the query pocket and $k$ values from $1$ to $300$, patch scores are computed and then summed to obtain a unique pocket score for each ligand type $F$ as follows: 
\begin{equation}
  \begin{split}
     \mbox{Pocket\_score}(P,F,k)\, = & \, \sum_{j=1}^{N_P} \mbox{Patch\_score}(p_j,F,k),\\   
  \end{split}
  \label{eqn:pocket_score}
\end{equation}
where $p_j$, $j=1,\ldots, N_P$ are the patches within the pocket $P$, and $N_P$ is the number of such patches. Thus, a pocket has a certain number of patches (see the last row of Table \ref{table:dataset}) and the score for a query pocket $P$ for a ligand type $F$ is computed as the sum of the score of each patch for the ligand $F$.

For each query pocket, we compute the pocket score for each of the $11$ ligand type in the database. The ligand with the highest Pocket\_score is predicted to bind to the query pocket. We then compare these $11$ scores and look at the largest one (Top 1) and the largest three (Top 3) to obtain the number of successful predictions (see Table \ref{table:prediction_accuracies}). For each ligand type, the number of successful cases is divided by the number of query pockets of that type in the pocket database to obtain prediction accuracies. For each prediction, the results are shown for the $k$ value which maximizes the average prediction accuracy. It turns out that the average prediction accuracy is maximized for small values of $k$ as shown in Table \ref{table:prediction_accuracies}.

When we look at the average prediction accuracies, $K_5$ performs best (with 41.0\% correct prediction) among the 3DKDs. For the individual ligand types, $K_3$ performs best only for FUC and HEM, while $K_4$ gives the highest accuracies for AMP, ATP, FMN, FUC, GLC, and MAN. Among the 3DKDs, $K_5$ is the one giving the best average prediction and highest prediction accuracies (including ties) for almost all ligand types except for AMP, GLC, and HEM. In Table \ref{table:confusionsK5}, we take a closer look at the results for $K_5$ by forming the confusion matrix with the true positives being along the diagonal. According to Table \ref{table:confusionsK5}, seven AMP queries are predicted as ATP, and similarly, five ATP queries are predicted as AMP. This may be excused due to similar shapes of AMP and ATP (only differing from each other by two phosphate groups.) Other similar shaped pairs can be observed among sugar molecules FUC, GAL, GLC, and MAN. FUC as a query gives one false positive (GLC), while GAL as a query is confused seven times with other sugar molecules (three times with FUC and four times with GLC). GLC as a query is confused five times with FUC, three times with GAL, and three times with MAN. Similarly, MAN as a query gives twelve false positives from the sugar group (FUC, nine times; GAL, two times; and GLC once). Thus, the confusion between ligands by 3DKD is quite reasonable, capturing similarity of binding pockets of similar ligand molecules. In the results for Top 3, $K_5$ is still the one showing the best average performance among the 3DKDs, and giving the highest accuracies (including ties) for almost all ligand types except FAD and GLC.

In Table \ref{table:prediction_accuracies}, we also provide a comparison of our approach with a former binding prediction method named Pocket-Surfer \cite{ChikhiSaelKihara2010}. In Pocket-Surfer, 3DZDs are used as shape descriptors of a pocket. In addition to shape information, the pocket size is also utilized in Pocket-Surfer as a classification measure by an optimal weighted average of scores from both shape and pocket size. In our work, we only use shape information from 3DKDs without employing the pocket size in the scoring functions. When Top 1 is the classification criterion, 3DKD predicts better for seven ligand types and the average prediction. With Top 3 classification, Pocket-Surfer performs better than 3DKD for nine (out of eleven) ligand types. We believe that this is partly due to the fact that 3DKD employs shape information only, whereas Pocket-Surfer also used pocket size, which is often critical to distinguish ligands of the different sizes. We also show results with random prediction, in which each query pocket is scored based on a randomly shuffled pocket database, averaged over $3000$ randomizations. It is clear that 3DKD outperforms the random prediction in all cases.

\begin{table}
	\centering
	\caption{Confusion Table for $K_5$}
    \label{table:confusionsK5}
	\footnotesize
		\begin{tabular}{|c|ccccccccccc|}
			\hline
            		&  \rotatebox{90}{AMP} & \rotatebox{90}{ATP} & \rotatebox{90}{FAD} & \rotatebox{90}{FMN} & \rotatebox{90}{FUC} & \rotatebox{90}{GAL} & \rotatebox{90}{GLC} & \rotatebox{90}{HEM} & \rotatebox{90}{MAN} & \rotatebox{90}{NAD} & \rotatebox{90}{PLM} \\
                    \hline
    			AMP  &  6 &    7 &    5 &    4 &    0 &    3 &    1 &    \textbf{9} &    1 &    5 &    3 \\
				ATP  &  5 &   \textbf{12} &    1 &    0 &    3 &    2 &    4 &    3 &    1 &    7 &    6 \\
     			FAD  &  1 &    6 &   \textbf{40} &    1 &    3 &    1 &    6 &   12 &    2 &    6 &    4 \\
     			FMN  &  3 &    1 &    5 &   \textbf{11} &    3 &    1 &    5 &    9 &    1 &    5 &    5 \\
     			FUC  &  0 &    0 &    0 &    0 &    \textbf{6} &    0 &    1 &    0 &    0 &    0 &    0 \\
     			GAL  &  0 &    1 &    1 &    1 &    3 &    \textbf{4} &    \textbf{4} &    0 &    0 &    0 &    1 \\
     			GLC  &  2 &    0 &    2 &    3 &    5 &    3 &    \textbf{6} &    1 &    3 &    2 &    0 \\
     			HEM  &  3 &    3 &    1 &    1 &    0 &    0 &    1 &  \textbf{127} &    0 &    0 &   10 \\
     			MAN  &  0 &    1 &    1 &    1 &    9 &    2 &    1 &    0 &   \textbf{16} &    1 &    1 \\
     			NAD  &  3 &    5 &    4 &    1 &    1 &    0 &    4 &    3 &    0 &   \textbf{16} &    2 \\
     			PLM  &  0 &    1 &    2 &    0 &    0 &    1 &    1 &    2 &    1 &    0 &    \textbf{3} \\
			\hline
		\end{tabular}
\end{table}


\begin{table}[!ht]
\renewcommand{\arraystretch}{1.3}
\centering
\begin{threeparttable}[b]
\caption{Computational Times for $K_5$ (seconds)}
\label{table:cpu_times}
\footnotesize
		\begin{tabular}{|c|c|c|}
        	\hline
					                  &    Laptop   &  PC       \\
			\hline
			
			
			Descriptors       &    0.5789   &  0.2733   \\
			\hline
			
			Pocket Score      &    0.4156   &  0.1112   \\
			\hline
			Search            & 3.5763e-05  &  1.0907e-05   \\
			\hline
		\end{tabular}
\end{threeparttable}
\end{table}


Table~\ref{table:cpu_times} shows the time took for computing 3DKD of order up to $5$ as outlined in steps (1)$-$(8) in Section~\ref{sec:Computation}. The programs were tested on two different platforms: a Windows laptop with i3 CPU of 2.53 GHz and 4 GB memory using MATLAB 2009b 64-bit, version 7.9, and on a Linux workstation with Xeon CPU of 3.60 GHz and 94 GB memory and using MATLAB 2016b 64-bit, version 9.1. On each computer, MATLAB is limited to a single computational thread to perform its jobs. The first row of Table~\ref{table:cpu_times} shows the average CPU times spent on computing 3DKD of order up to $5$, where the average is taken among $1608$ local protein surface patches from Test I. The computation of 3DKDs can be done in half a second in a laptop, whereas it finishes twice faster on the PC. In the second row of Table~\ref{table:cpu_times}, we also show the computational time from the binding prediction test, assuming that the 3DKDs of all 11,039 patches in the patch database are precomputed and stored. Given a typical query pocket, the time it takes to compute the pocket scores in (\ref{eqn:pocket_score}) for $11$ ligand types and $k=1,2,\ldots, 10$ is about $0.4$ second in a laptop, and $3.7$ times faster in the second platform.



\section{Conclusion}
\label{sec:Conc}

In this paper, we have developed a novel set of local descriptors, three-dimensional Krawtchouk descriptors (3DKD), for identification and comparison of local regions of 3D voxelized oriented surfaces. Our approach is based on 3D Krawtchouk moments. While obtaining the rotation, size, and position independent invariants, we have preserved the critical ability of Krawtchouk moments to extract local features of a 3D surface from any region-of-interest. The locality property is due to the 3D weight function given in the definition of Krawtchouk polynomials. The weight contains three parameters $p_x$, $p_y$, and $p_z$, shifting the center of the local surface region along the $x$, $y$, and $z$-axes, respectively. We have noticed that, for each triplet $(p_x,p_y,p_z)$, the coverage of the weight function is different, which prevents Krawtchouk moments from being translation and rotation invariant. To overcome these problems, we have computed the 3D weight function corresponding to $(0.5,0.5,0.5)$ (center of the 3D grid) and used it for other local regions by translating the graph of the weight function as needed. To achieve rotational invariance, we utilized the surface normal at the vertex of the local surface region and computed the eigenvectors of the local inertia matrix. Among the eight possible different orientations, we have chosen the one positioning the vertex normal in a fixed octant in 3D space. We have also provided a detailed scheme for efficient computation of 3D Krawtchouk descriptors. 

We have tested the discriminative performances of 3DKD on three test problems. For each test, we have used $K_3$, $K_4$, and $K_5$, namely 3DKD of the order up to $3$, $4$, and $5$, respectively. In the first test, the results have been comparable, while $K_5$ has the best recognition accuracies for predicting the top match. The second test demonstrated that $K_5$, among the 3DKD, is the most robust set of descriptors to small changes in patch location. We have also compared 3DKD with 3DZD. 3DKD shows much better recognition accuracies than 3DZD in all cases reported. As the third test, we have employed 3DKD for prediction of ligand binding sites on protein surfaces. 3DKD showed better performance than Patch-Surfer when Top 1 prediction was considered. From the results of the second and the third tests suggest that 3DKD is more sensitive than 3DZD in detecting subtle shape similarity. The results on the binding ligand prediction were obtained by only considering geometric shape information of protein surface. Therefore further improvement is expected by integrating other features, such as the electrostatic potential or other physicochemical properties, to characterize protein surface regions, which is analogous to representing color images rather than black-and-white images of protein surfaces.


\section*{Acknowledgments}
This work was supported by the National Science Foundation (DMS1614777 and DMS1614661). DK also acknowledges support from National Institutes of Health, (R01GM123055) and the National Science Foundation (CMMI1825941) and the Purdue Institute for Drug Discovery. The authors would like to thank Xiaolei Zhu for the help in voxelizing protein surfaces and Juan Esquivel-Rodriguez for the help in running LZerD software.






\bigskip
\bigskip

\noindent\textbf{Atilla Sit} is currently an Assistant Professor with the Department of Mathematics and Statistics, Eastern Kentucky University, Richmond, KY, USA. He received the B.S. degree from Middle East Technical University, Ankara, Turkey, the M.S. degree from Bogazici University, Istanbul, Turkey, and the Ph.D. degree from Iowa State University, Ames, IA, USA, in 2001, 2005, and 2010, respectively. His current research interests include image/structure analysis, orthogonal systems, special functions, protein structure determination and comparison, and protein classification.

\bigskip

\noindent\textbf{Daisuke Kihara} is currently a Professor with the Department of Biological Sciences and the Department of Computer Science, Purdue University, West Lafayette, IN, USA. He received the B.S. degree from the University of Tokyo, Japan, in 1994, and M.S. and Ph.D. degrees from Kyoto University, Japan, in 1996 and 1999, respectively. His research projects include biomolecular shape comparison, computational drug design, protein tertiary structure prediction, protein-protein docking, and other geometrical problems in computational biology. He has authored over 150 research papers and book chapters. He is named Showalter University Faculty Scholar from Purdue University in 2013.

\end{document}